%% file: main.tex
\documentclass[acmlarge]{acmart}

\acmJournal{IMWUT}
\acmVolume{1}
\acmNumber{1}
\acmArticle{1}
\acmYear{2024}
\acmMonth{1}

\usepackage{xspace}
\usepackage[utf8]{inputenc}
\usepackage{enumitem}
\usepackage{xcolor}
\usepackage{mdframed}
\usepackage{subcaption}
\usepackage{fancyhdr}
\usepackage{multirow}

\mdfsetup{skipabove=1.5em,skipbelow=1.5em}

\newcommand{\IAC}[1]{\textcolor{black}{#1}} 
\newcommand{\IIAC}[1]{\textcolor{black}{#1}}
\newcommand{\RI}[1]{\textcolor{black}{#1}}
\newcommand{\RII}[1]{\textcolor{black}{#1}}

\begin{document}

\newpage

\title{PPG as a Bridge: Cross-Device Authentication for Smart Wearables with Photoplethysmography}
\author{Jiacheng Liu}
\authornote{Co-first authors}
\email{jl4596@cornell.edu}
\orcid{0009-0007-4525-0352}
\affiliation{%
  \institution{Cornell University}
  \state{Ithaca}
  \country{USA}
}

\author{Jiankai Tang}
\authornotemark[1]
\orcid{0009-0009-5388-4552}
\email{tjk24@mails.tsinghua.edu.cn}
\affiliation{%
  \institution{Tsinghua University}
  \state{Beijing}
  \country{China}
}

\author{Guangye Zhao}
\orcid{0009-0006-7616-6092}
\email{zhaogy23@mails.tsinghua.edu.cn}
\affiliation{%
  \institution{Tsinghua University}
  \state{Beijing}
  \country{China}
}

\author{Ruichen Gui}
\orcid{0009-0009-9496-9106}
\email{ruichg@uw.edu}
\affiliation{%
  \institution{University of Washington}
  \city{Seattle}
  \country{USA}
}

\author{Songqin Cheng}
\orcid{0009-0002-3564-5741}
\email{csq24@mails.tsinghua.edu.cn}
\affiliation{%
  \institution{Tsinghua University}
  \state{Beijing}
  \country{China}
}

\author{Taiting Lu}
\affiliation{%
  \institution{Pennsylvania State University}
  \city{University Park}
  \state{PA}
  \country{USA}
}

\author{Jian Liu}
\affiliation{%
  \institution{Ant Group}
  \city{Hangzhou}
  \country{China}
}

\author{Weiqiang Wang}
\affiliation{%
  \institution{Ant Group}
  \city{Hangzhou}
  \country{China}
}

\author{Mahanth Gowda}
\affiliation{%
  \institution{Pennsylvania State University}
  \city{University Park}
  \state{PA}
  \country{USA}
}

\author{Yuanchun Shi}
\orcid{0000-0003-2273-6927}
\email{shiyc@tsinghua.edu.cn}
\affiliation{%
  \institution{Tsinghua University}
  \state{Beijing}
  \country{China}
}

\author{Yuntao Wang}
\authornote{Corresponding author.}
\email{yuntaowang@tsinghua.edu.cn}
\orcid{0000-0002-4249-8893}
\affiliation{%
  \institution{Tsinghua University}
  \state{Beijing}
  \country{China}
}

\renewcommand{\shortauthors}{Trovato et al.}

\begin{abstract}



As smart wearable devices become increasingly powerful and pervasive, protecting user privacy on these devices has emerged as a critical challenge. While existing authentication mechanisms are available for interaction-rich devices such as smartwatches, enabling on-device authentication (ODA) on interaction-limited wearables—including rings, earphones, glasses, and wristbands—remains difficult. Moreover, as users increasingly own multiple smart devices, relying on device-specific authentication methods becomes redundant and burdensome. To address these challenges, we present PPGTransID, a ubiquitous and unobtrusive cross-device authentication (CDA) approach that leverages the real-time physiological consistency of photoplethysmography (PPG) signals across the human body. PPGTransID utilizes the widely available PPG sensors on wearable devices to capture users’ physiological signals and compares them with remote PPG (rPPG) signals extracted from a smartphone camera, where robust face-based authentication is already established. In doing so, PPGTransID securely transfers the reliable authentication status of the smartphone to nearby wearable devices without requiring additional user interaction. An evaluation (N=33) shows that PPGTransID achieves a balanced accuracy of 95.5\% and generalizes across multiple wearable form factors. Robustness experiments (N=10) demonstrate resilience to variations in lighting, camera placement, and user behavior, while a real-time usability study (N=14) confirms reliable performance with minimal interaction burden.

\end{abstract}

\ccsdesc[300]{Human-centered computing~Ubiquitous and mobile computing systems and tools}

\keywords{Authentication, Wearable Device, Smart Sensing, Photoplethysmography}
\begin{teaserfigure}
  \includegraphics[width=\textwidth]{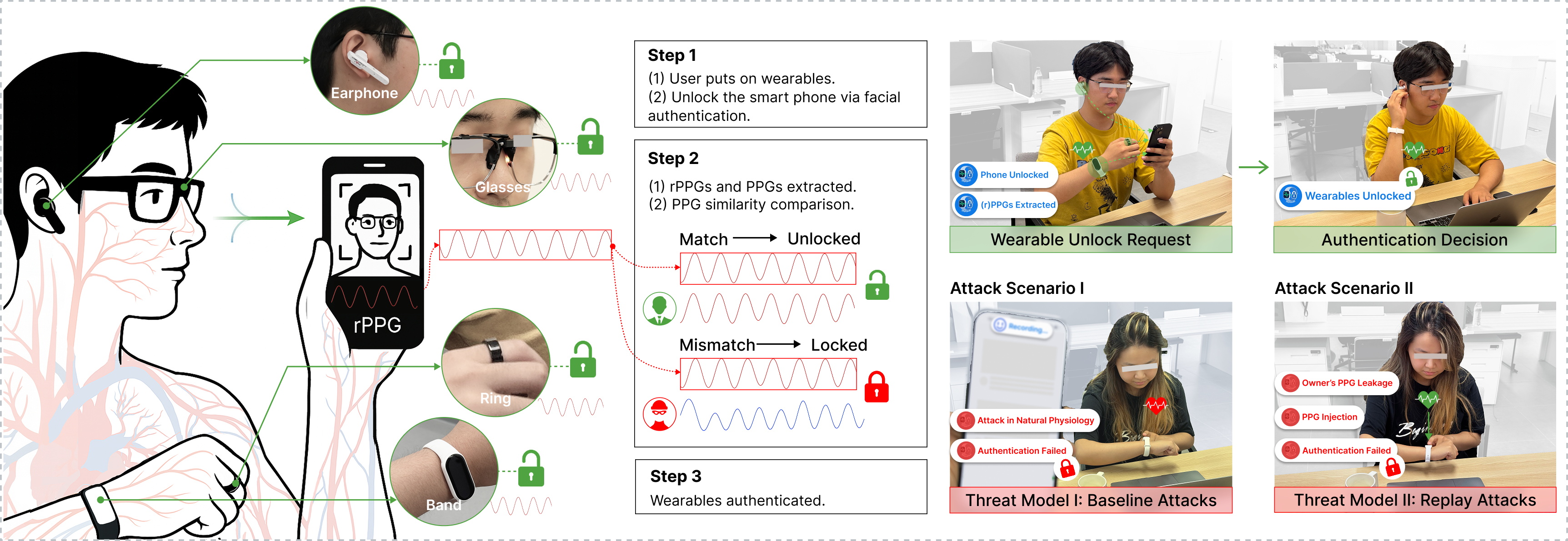}
  \caption{\textbf{System Overview of PPGTransID.} PPGTransID is a cross-device authentication method that leverages the physiological consistency between rPPG signals captured by smartphones and PPG signals from smart wearables to transfer authentication status from the smartphone to co-worn devices. By doing so, it enables ubiquitous and unobtrusive authentication, allowing users to authenticate multiple wearables during everyday smartphone use with minimal disruption.}
  \Description{This image illustrates the PPGTransID system, which enables secure cross-device authentication from smartphones to wearable devices using synchronized physiological signals. The process involves unlocking the owner's smartphone, video-based rPPG extraction, and authentication by matching signals. The system supports various wearables, such as earphones, glasses, rings, and bands. It also demonstrates two attack scenarios, baseline and replay attacks, and how the system resists unauthorized access by comparing real-time physiological data across devices.}
  \label{fig:teaser}
\end{teaserfigure}

\received{20 February 2007}
\received[revised]{12 March 2009}
\received[accepted]{5 June 2009}

\maketitle

\section{INTRODUCTION} \label{chap:intro}
Identity authentication is a fundamental requirement for the broader adoption of smart wearables in healthcare, communication, and payment. \RII{However, traditional \textbf{On-device Authentication (ODA)} — which typically requires template enrollment and on-device verification — can be difficult to deploy on wearables with constrained computation, sensing, and interaction. A promising alternative is \textbf{Cross-device Authentication (CDA)~\cite{holz2016demand}}, where identity verified on a trusted token device~\cite{findling2016shakeunlock} is seamlessly transferred to another device. Security is maintained by verifying that the token device and wearables are co-present on the same user. In other words, once the token device has authenticated the user through mature methods such as FaceID or fingerprint recognition, wearables only need to confirm shared user presence rather than perform full authentication themselves. A practical example is the Apple Watch: When users unlock their iPhone, the paired Apple Watch automatically unlocks without requiring any additional interaction.}

\RII{Compared to traditional ODA, CDA eliminates per-device template enrollment and relies on real-time device-to-device communication to confirm co-presence.} Consequently, CDA inherently offers stronger long-term reliability, avoiding the degradation faced by ODA systems due to changes in physiological or behavioral characteristics~\cite{hintze2015cormorant}. \RII{It also shifts the authentication burden from individual wearables to the token device, reducing the constraints imposed by wearables' limited interaction and standalone authentication capabilities.}

However, existing CDA methods suffer from several inherent limitations. The first limitation lies in the choice of the token device. Prior research~\cite{wu2022use,findling2016shakeunlock} has not adequately justified the reliability or suitability of this source device. For instance, if a wristband is designated to unlock other devices, a feasibility issue arises when the band itself is difficult to unlock. \IAC{Second, a gap remains in CDA methods that generalize across diverse wearables. Some approaches rely on IMU or motion data~\cite{wu2022use,findling2016shakeunlock, mare2018saw, huang2019id}, which can be reliable but are limited to certain wearables.} In practical use, most wearable devices do not require specific motion patterns, nor do they frequently move in coordination with the trusted token device. This introduces unnecessary interaction overhead and deviates from natural user behavior. \IAC{Third, systems that depend on proprietary device-level trust or manual user confirmation—such as Apple Watch’s proximity-based unlocking~\cite{Ranganathan2017AreWR}—are difficult to extend across brands, and manual confirmation adds user burden and potential security risks.}



In this paper, we present PPGTransID, a PPG-based CDA method that enables ubiquitous and unobtrusive authentication for smart wearables. As shown in Figure~\ref{fig:teaser}, after a user initiates facial authentication to unlock their smartphone, the phone automatically captures a short facial video in the background while simultaneously collecting synchronized PPG signals from optical sensors embedded in co-worn wearables. PPGTransID extracts remote photoplethysmography (rPPG) from the facial video and correlates it with wearable PPG signals to make authentication decisions. Once verified, the authentication state can be maintained through mechanisms such as wear detection~\cite{kheirkhahan2017power} or continuous heart rate monitoring~\cite{armanac2022reliability}. In essence, PPGTransID allows users to authenticate all of their smart wearables at once with minimal disruption to everyday smartphone use.


PPGTransID is designed to address the limitations of existing methods. First, smartphones have established mature unlocking mechanisms with strong privacy attributes, making them a reasonable choice as token devices~\cite{schroff2015facenet}. Second, smartphones are frequently used in daily life and can acquire rPPG signals from the front camera without introducing additional user interactions~\cite{tang2024camerabased}. Meanwhile, most wearables are already equipped with optical sensors that allow unobtrusive and implicit acquisition of PPG signals once worn~\cite{tang2025dataset}. Third, the method relies on the consistency of real-time physiological features across different body sites rather than device-level trust, making it less susceptible to proximity attacks. 




To evaluate the \textbf{ubiquity and usability} of PPGTransID, we posed the following research questions.
\begin{itemize}
    \item RQ1: Is PPGTransID applicable across various common smart wearables?
    \item RQ2: Can PPGTransID effectively resist diverse attacks against CDA?
    \item \IAC{RQ3: Can PPGTransID remain robust under diverse environmental conditions and user behaviors?}
    \item RQ4: Does PPGTransID provide a favorable user experience?
\end{itemize}


To address these questions, we conducted a user study with 33 participants to evaluate the system on three custom-built wearable devices (band, glasses, ring) and a commercial earphone, under both sitting and standing postures. With an XGBoost classifier and our preprocessing pipeline, the system achieved a balanced accuracy (BAC) of 95.5\% without user calibration, while maintaining efficient memory usage (0.46 M parameters) and low inference latency (9.3 ms). \RII{The system also demonstrated robust performance across wearables, body postures, rPPG extraction methods, and environmental conditions, and exhibited resistance to replay attacks. Additionally, we explored potential use cases by evaluating performance on different token devices, unseen wearable devices, and shorter usage durations.}

\RII{To further evaluate the system's robustness to variations in lighting, head motion, and camera positioning, a user study involving 10 participants was conducted. The results indicate that these condition changes introduce no significant impact on the reliability of the approach.}

To examine real-world applicability, we implemented a real-time demonstration of PPGTransID on an iPhone 15 and a laptop. A study with 14 participants simulated two scenarios: baseline attacks and authentication while scrolling the screen. The system achieved consistent BACs of 97.7\% and 96.7\%, respectively. Participants also reported favorable perceptions in the system’s usability and future applications. Taken together, these results are the major contributions of this paper:



\begin{itemize}
    \item Introduced PPGTransID, a ubiquitous and unobtrusive cross-device authentication approach that exploits the real-time physiological consistency between (r)PPG captured by the token device and wearable PPG signals for smart wearable authentication.
    \item Conducted comprehensive user studies to evaluate PPGTransID with respect to its resistance to defined threat models, ubiquity across diverse wearable devices, and robustness under non-ideal conditions.
    \item Developed a real-time implementation of PPGTransID on an iPhone and a laptop, and validated its usability and acceptance through interactive demonstrations.
\end{itemize}

\section{RELATED WORK}

\subsection{Cross-device Authentication}


CDA has been investigated in prior work as a way to extend authentication beyond a single device by securely transferring identity from one trusted device to others. Early work showed that signals observable across devices can provide such a basis. Examples include ambient RF noise~\cite{Jin2020HarnessingTA}, hand resonance~\cite{Wang2016TouchandguardSP}, bodily electrical response~\cite{Wang2022EnablingST}, bone conduction~\cite{Wang2025S2PairSA}, and heartbeat-based cross-modal keys~\cite{Wei2024FaceFingerEV}. These studies demonstrated the feasibility of shared physical or physiological features for secure transfer, though most remained at the pairing or key-generation level.  


More explicit forms of CDA leverage human motion to propagate authentication across devices. ShakeUnlock~\cite{findling2016shakeunlock} employed synchronized shaking to transfer an unlocked status from a smartwatch to a smartphone. SAW~\cite{mare2018saw} enabled wristband-based authentication for desktop computers by matching motion patterns between wearable sensors and workstation interactions. Related work has also explored authentication transfer across heterogeneous systems using motion signatures captured during daily activities. For example, No-Need-to-Shake-It~\cite{wu2022use} advanced implicit authentication by correlating motion patterns between a wristband and everyday objects equipped with inertial sensors, while Au-Id~\cite{huang2019id} leveraged motion features extracted from sequential human activities using RFID for automatic user identification and authentication. These systems demonstrate that body motion can act as a transferable biometric across devices and sensing modalities. However, such approaches typically rely on explicit user movements, activity-dependent motion patterns, or additional sensing infrastructure, which limits their ubiquity and suitability for unobtrusive authentication. 

Moreover, commercial systems, such as iPhone unlocking Apple Watch, rely on proximity-based transfer after FaceID, though such assumptions can be vulnerable~\cite{Ranganathan2017AreWR}. In contrast, our work departs from these paradigms by leveraging PPG—a sensing modality already embedded in smartphones and wearables—to enable ubiquitous and unobtrusive CDA without requiring explicit gestures, additional sensors, or proximity assumptions


\subsection{Remote Photoplethysmography Algorithms}

\RII{The advancement of rPPG now enables smartphones and other devices with mature authentication to serve as reliable token devices for PPG-based CDA.} Traditional rPPG algorithms primarily rely on signal processing and color space analysis to extract blood volume pulse from facial videos. The seminal work of \citet{verkruysse2008remote} demonstrated that plethysmographic signals can be captured remotely using ambient light and RGB cameras, establishing the foundation for this field. ICA-based methods such as \citet{poh2010advancements} improved robustness by separating pulse signals from noise, while CHROM~\cite{de2013robust} enhanced color space projections for more stable heart rate estimation. To mitigate motion artifacts, the POS algorithm~\cite{wang2016algorithmic} introduced a projection onto a plane orthogonal to the skin tone, significantly improving robustness under movement. More recently, Face2PPG~\cite{casado2023face2ppg} proposed an unsupervised framework that extracts blood volume pulse signals without explicit supervision, highlighting progress toward real-world deployment. Despite these advances, most traditional methods assume steady illumination and limited motion, restricting their reliability in unconstrained environments.

Deep learning methods have been developed to overcome these limitations by learning spatio-temporal representations directly from video. Early neural approaches such as DeepPhys~\cite{chen2018deepphys} and PhysNet~\cite{yu2019remote, liu2024summit} demonstrated the feasibility of end-to-end rPPG estimation. Transformer-based architectures including PhysFormer~\cite{physformer,tang2024spikingphysformer} and RhythmFormer~\cite{zou2025rhythmformer} further improved accuracy by modeling long-range temporal dependencies. More recently, PhysMamba~\cite{luo2024physmamba} and MaKAN-Mixer~\cite{zhang2025makan} advanced robustness by combining state-space or channel-mixing modules with temporal dynamics amplification. However, the majority of these methods are validated only on high-quality stationary videos, limiting applicability to everyday mobile use. In contrast, we deploy TN-rPPG~\cite{wang2024plug}, which leverages temporal neighborhood aggregation to enable reliable heart rate extraction from handheld smartphone videos with strong motion noise, and ME-rPPG~\cite{wang2025memory}, which provides real-time PPG waveform estimation for improved user experience in continuous monitoring scenarios.

\subsection{Authentication on Smart Wearables}

Wearable authentication has been studied across multiple device form factors, ranging from wristbands to earbuds and smart glasses. PPG-based approaches have been proposed for on-wrist wearable authentication, where commercial devices such as Samsung Gear and Empatica E4 have demonstrated continuous implicit authentication in daily life~\cite{ekiz2020can}. Building on this, PPGPass~\cite{cao2020ppgpass} employed dual-channel PPG with adaptive filtering to improve robustness, while M-PPG~\cite{pan2024m} introduced motor-induced vibrations to enhance signal distinctiveness, though all remain tied to per-device enrollment. Moving from the wrist to the ear, EarPass~\cite{li2023earpass} explored in-ear PPG with a quality assessment for motion artifacts, and EarID~\cite{10.1145/3706599.3719788} advanced ear canal sensing, achieving 98.1\% balanced accuracy but requiring specialized sensors. Extending further to head-mounted devices, SkullID~\cite{10.1145/3613904.3642506} leveraged through-skull sound conduction for smartglasses, achieving error rates below 3\% with robustness to donning variations, yet still confined to glasses-specific acoustic hardware. 

These studies reveal the rich potential of physiological and anatomical signals for wearable authentication, but they share fundamental limitations: device-specific design, template-based registration, and limited generalizability across heterogeneous devices. In contrast, our work leverages PPG consistency between smartphones and wearables to transfer identity across devices, eliminating per-device enrollment while ensuring long-term reliability and low-burden user experience.

\section{SYSTEM OVERVIEW}


\subsection{System Architecture}

As illustrated in Figure~\ref{fig:teaser}, PPGTransID consists of three key components:
(1) smart wearables equipped with PPG sensors,
(2) a mobile phone registered to the legitimate user with mature facial authentication capability, and
(3) a CDA process running on the smartphone or an external server.

The CDA process is triggered when the user unlocks the phone via facial authentication. During this process, the frontal camera captures a short facial video, which is used to extract rPPG signals for CDA. Since reliable PPG-based authentication requires a longer signal duration than conventional facial authentication, the camera remains active for several additional seconds to acquire rPPG signals of sufficient length. Meanwhile, synchronized PPG signals are collected from the embedded sensors on the wearable devices. The CDA process then preprocesses both signals and computes a correlation score between the rPPG and wearable PPG signals to verify device ownership.

In this work, we focus on four common types of wearables: smart earphones, smart glasses, smart rings, and smart bands. Unlike mobile phones and smartwatches, these devices typically have limited input interfaces, sensing capabilities, and immature authentication mechanisms. PPGTransID is designed to address this challenge by enabling a \textbf{ubiquitous, form-factor-agnostic authentication mechanism across heterogeneous wearable devices}.


\subsection{Theory of Operation}\label{sec:theory}



PPG, as a widely adopted and commercially mature sensing modality, naturally satisfies the key requirements of a unified and form-factor-agnostic CDA mechanism:
(1) it remains consistent across different devices worn by the same user,
(2) it preserves sufficient inter-user distinctiveness for reliable authentication, and
(3) it exhibits transient, moment-to-moment variations that mitigate the risk of misuse from leaked or replayed signals.




\textbf{Cross-device consistency.} PPG is an optical signal that reflects changes in light absorption and scattering induced by cardiac-driven pulsatile blood flow. Although differences in hardware characteristics and local tissue properties introduce morphological variations and small temporal offsets in PPG waveforms captured at different body locations, synchronized cross-device PPG signals still exhibit high correlation. Prior studies have shown that these signals share common physiological features, such as heart rate and pulse rate variability, across devices worn by the same individual~\cite{block2020conventional, nitzan2001difference, yuda2020differences, burma2024heart}.



\textbf{Inter-user distinctiveness.} PPG features exhibit substantial inter-user variability arising from physiological factors such as age, medical conditions, sex, and body mass index, which provides a natural basis for user discrimination~\cite{avram2019real, natarajan2020heart, park2022photoplethysmogram}. Using the data collected in Study~1 (Section~\ref{chap:study1}), we observe that synchronized cross-device PPG signals from the same user demonstrate significantly higher similarity than those from different users. Specifically, synchronized intra-user PPG pairs achieve a higher Pearson correlation (mean = 0.830, SD = 0.070) and coherence similarity (mean = 0.743, SD = 0.071). In contrast, inter-user PPG pairs exhibit substantially lower correlation (mean = 0.501, SD = 0.007) and coherence similarity (mean = 0.326, SD = 0.053).




\textbf{Temporal variations.} Without the need for template registration, CDA approaches typically rely on transient, momentary features shared across devices to enhance security, such as IMU signals during device shaking~\cite{findling2016shakeunlock} or patterns induced by human–object interaction~\cite{wu2022use}. While prior work on PPG-based ODA has demonstrated that PPG signals contain stable morphological characteristics over time~\cite{tang2025exploring, cao2020ppgpass, zhao2022robust, pan2024m, wan2024deep}, these signals also exhibit natural temporal variations driven by dynamic cardiac activity. This combination of temporal stability and short-term variability enables PPG to support secure CDA without explicit template registration. We further analyze this property in Section~\ref{sec:feature extraction} and Section~\ref{sec:threat model II}.




\subsection{Threat Model}\label{sec:threat model}


We consider an attacker A who has illegally obtained a wearable device W and aims to deceive the authentication system into believing that a legitimate user L is wearing and using W. In this work, we examine two threat models.

\textbf{Baseline attacks (Threat Model I).} A wears the stolen W and attempts to authenticate using their own physiological signals. Specifically, A wears W while the L unlocks their mobile phone and initiates the CDA process. This threat model evaluates whether the system can be fooled by cross-device signals originating from different users.

\textbf{Replay attacks (Threat Model II).} A obtains recorded PPG signals of L through unintended leakage or external monitoring. These recorded signals are subsequently injected into the stolen W while L unlocks the mobile phone and initiates the CDA process. This threat model assesses whether the system can be deceived using a legitimate user’s historical physiological data, a widely recognized risk in authentication systems that rely on biometric signals~\cite{shin2024skullid}.

\textbf{Assumptions and Non-considered Scenarios.} Mimicry attacks, in which an attacker attempts to imitate a victim’s motion patterns or physiological states, have been discussed in prior CDA studies~\cite{wu2022use, findling2016shakeunlock, shin2024skullid}. In our work, we assume that attackers cannot precisely and voluntarily control PPG signals, as PPG reflects involuntary cardiac activity. A potential strategy for partial mimicry is heart rate adaptation. To examine this possibility, we conducted an experiment with the participants in Study~3 (Section~\ref{chap: study 3}), where participants were instructed to align their heart rates with those of an experimenter using techniques such as deep breathing or light physical activity. The results showed no statistically significant reduction in heart rate differences after adaptation. Participants reported difficulty in deliberately lowering their heart rates (P7, P8), while others noted that post-exercise heart rates often exceeded the intended targets (P2, P5), making accurate physiological mimicry impractical. We therefore do not consider precise PPG mimicry capabilities in our threat model.

Some CDA systems require explicit user confirmation, such as notification-based approval, to authorize device access. While effective, such mechanisms represent an orthogonal engineering solution that can be incorporated into a wide range of CDA approaches, rather than a property intrinsic to the CDA method itself. Moreover, explicit user confirmation may introduce additional interaction overhead, particularly when multiple devices need to be unlocked. As this work focuses on signal-based CDA without explicit user intervention, we do not consider such approaches.

\subsection{Hardware Implementation}\label{sec:hardware}
\begin{figure}[t]
    \centering
    \includegraphics[width=0.7\linewidth]{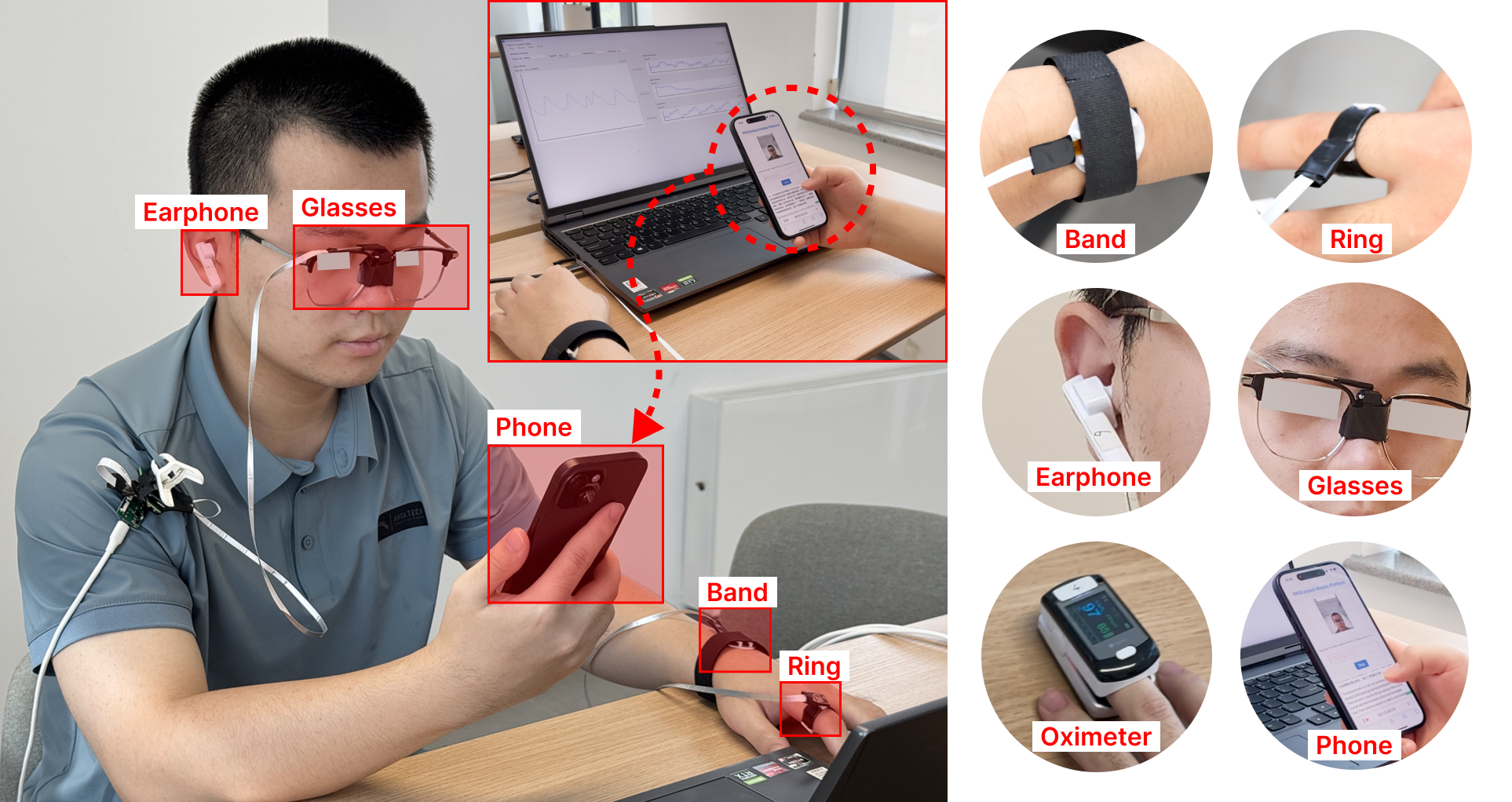}
    \caption{\textbf{Demonstration of Hardware Devices.} Illustration of device wearable placement and usage (left) and close-up views of the hardware used in our studies (right).}
    \Description{This image shows a participant using multiple wearable devices and a smartphone to demonstrate the PPGTransID system. The left side displays the experimental setup, including a band, ring, earphone, and phone, with data being monitored on a laptop. The right side highlights close-up views of each device used for physiological signal collection. The research focuses on integrating various hardware to enable secure cross-device identity verification using PPG signals.}
    \label{fig:wearable_setup}
\end{figure}

We developed three custom wearable devices (a ring, a band, and glasses), along with a commercial TH2318 PPG earphone, to support the study objectives. Additional equipment included a commercial CMS50E+ pulse oximeter, an OPPO A52 smartphone (for earphone data recording), an OPPO Find X3 Pro (for fingertip video capture), an iPhone 15 (for facial video capture), and an ASUS ROG Strix G15 G513QM laptop (for recording PPG signals from the custom wearables).

All custom devices shared the same electronic components but differed in mechanical design and configuration to optimize signal quality at their respective measurement sites. Each device employed an Analog Devices MAX30101 optical sensor, commonly used in prior work~\cite{punay2024extraction, iqbal2024wearable}. To accommodate variations in skin and vascular properties across different body sites, LED drive currents were individually tuned to ensure sufficient signal amplitude. The sensors were driven by an nRF52832 microcontroller, which streamed PPG signals via USB to a laptop at 110 samples per second.

The custom devices were encased in 3D-printed, skin-safe thermoplastic polyurethane (TPU) enclosures designed to fit the anatomical contours of the measurement sites and maintain stable sensor–skin contact. As shown in Figure~\ref{fig:wearable_setup}, the ring placed the sensor on the proximal phalanx of the index finger, the band positioned it on the dorsal forearm, and the glasses located it on the lateral side of the nasal bridge — the most commonly adopted site for PPG monitoring in glasses form factors~\cite{scandelli2024study, constant2015pulse}.

\begin{figure*}[t]
    \centering
    \includegraphics[width=\linewidth]{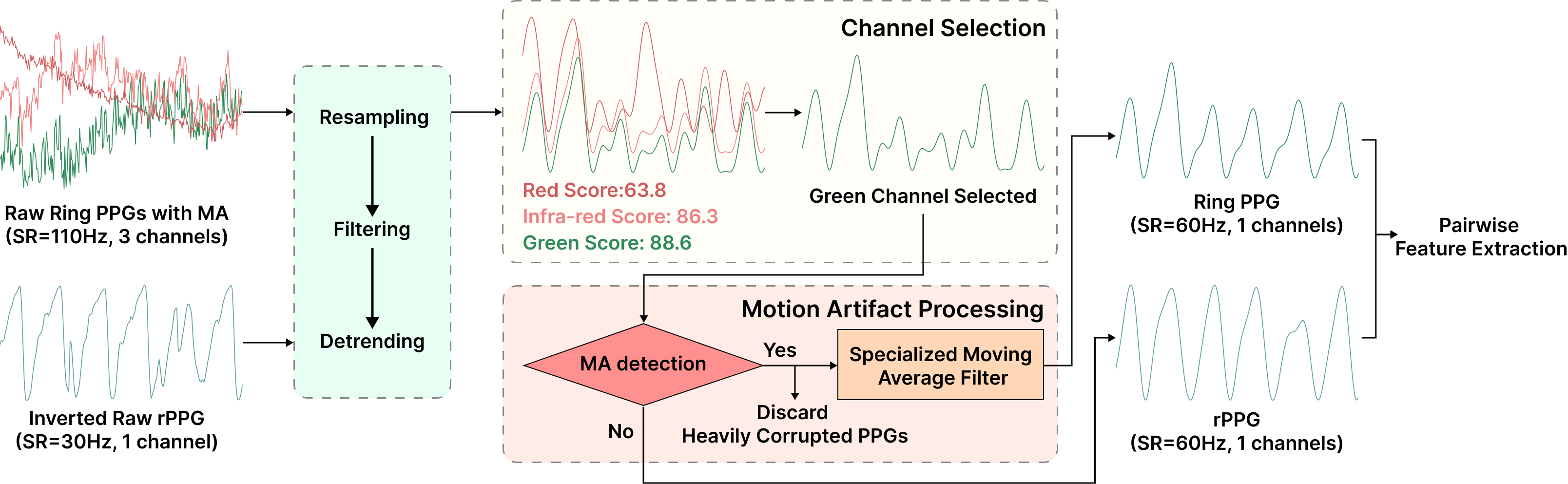}
    \caption{\textbf{Preprocessing Pipeline.} To normalize PPG signals across devices and sensing modalities, signals are resampled, filtered, and detrended. For multi-channel devices, a quality score is computed for each channel, and the best channel is selected. MA processing classifies signals as clean, weakly corrupted, or heavily corrupted; heavily corrupted signals are discarded, while weakly corrupted signals are mitigated using a specialized moving average filter. Pairwise features are then extracted from the cleaned PPG pairs.}
    \Description{The diagram illustrates the preprocessing pipeline for PPG signals collected from heterogeneous devices. Raw PPG signals are first resampled, filtered, and detrended to normalize temporal characteristics. For devices with multiple channels, a quality score is computed for each channel and the highest-quality channel is selected. The signals then undergo motion artifact processing, where clean signals are retained, weakly corrupted signals are mitigated using a moving average filter, and heavily corrupted signals are discarded. Finally, pairwise features are extracted from the cleaned PPG signal pairs for subsequent analysis.}
    \label{fig:system pipeline}
\end{figure*}

\section{SYSTEM PIPELINE}

PPG signals extracted from heterogeneous devices and sensing modalities differ in sampling rates, available signal channels, and susceptibility to motion artifacts (MA). To address these variations and enable reliable pairwise feature extraction, we propose a unified preprocessing pipeline that mitigates cross-source discrepancies among PPG signals and produces consistent pairwise features (Figure~\ref{fig:system pipeline}).

\subsection{rPPG Extraction}\label{sec:rppg}


We employed two complementary approaches to extract rPPG signals from facial videos. The first is a temporal normalization–based rPPG method (TN-rPPG~\cite{wang2024plug}), which processes 160 consecutive frames to generate a continuous waveform. By leveraging long temporal context, it is robust to subtle head motion and hand tremors, making it suitable for benchmarking authentication accuracy in controlled settings.

To support real-time interaction, we additionally used a memory-efficient single-frame inference method (ME-rPPG~\cite{wang2025memory}), which estimates rPPG signals in a streaming, frame-by-frame manner. Although initially less stable, its state-space design adapts over time and enables responsive real-time use in usability experiments.

Both methods follow a shared pipeline consisting of facial video capture, region-of-interest detection and cropping, frame preprocessing ($36 \times 36$), model inference, and waveform output. For clarity, we refer to TN-rPPG as Video-based rPPG and ME-rPPG as Frame-based rPPG throughout the paper.

As both models are trained using an MSE loss to reconstruct transmission-mode PPG captured by a clip-based sensor, whose polarity is opposite to that of reflectance PPG from wearables~\cite{ryals2023photoplethysmography}, we invert the rPPG signals before further processing.






\subsection{Cross-Device Signal Normalization} 

Different sensors, devices, and wearing positions lead to variations in PPG signal-to-noise ratio (SNR), waveform morphology, available channels, and sampling rates. To mitigate these differences, we apply a Butterworth bandpass filter (0.5–2 Hz) to all PPG signals, covering the typical resting heart rate range (48–100 bpm). All signals are then resampled to 60 Hz, segmented into 12-second windows, and detrended using the DC-removal method proposed by Hung et al.~\cite{hung2025reliable}.

As different wearable devices provide different numbers and types of PPG channels, we perform channel selection to select the channel with the best signal quality and unify the number of channels across devices. Prior studies have shown that skewness~\cite{zhao2022robust}, kurtosis~\cite{lambert2024impact,cao2020ppgpass}, relative spectral power~\cite{7605499}, and template matching~\cite{lim2018adaptive,6862843} are useful indicators for MA detection. However, we observe that the distributions of these metrics vary substantially across devices, making it difficult to define a single device-agnostic threshold.

To address this challenge, we compute a unified channel quality score for each PPG channel as:

\begin{equation}
    \mathrm{S}
    =
    \lambda_{S}\,\mathbb{I}(-0.5 \le \mathrm{S} \le 0.8)
    +
    \lambda_{K}\,\mathbb{I}(\mathrm{K} \le 0.7)
    +
    \lambda_{R}\,\mathrm{R}
    +
    \lambda_{T}\,\mathrm{T}
\end{equation}

where $\mathbb{I}(\cdot)$ is an indicator function that equals 1 if the condition holds and 0 otherwise. We treat skewness and kurtosis as binary indicators, as long as they fall within empirically reasonable ranges, since their exact values do not reliably reflect signal quality across devices. Based on empirical observations, we set $\lambda_{S}=\lambda_{K}=0.1$, and $\lambda_{R}=\lambda_{T}=0.4$. The channel with the highest score is selected for subsequent processing.

\subsection{Motion Artifact Processing}\label{sec:MA processing}

\begin{figure}[t]
  \centering
  \includegraphics[width=0.5\columnwidth]{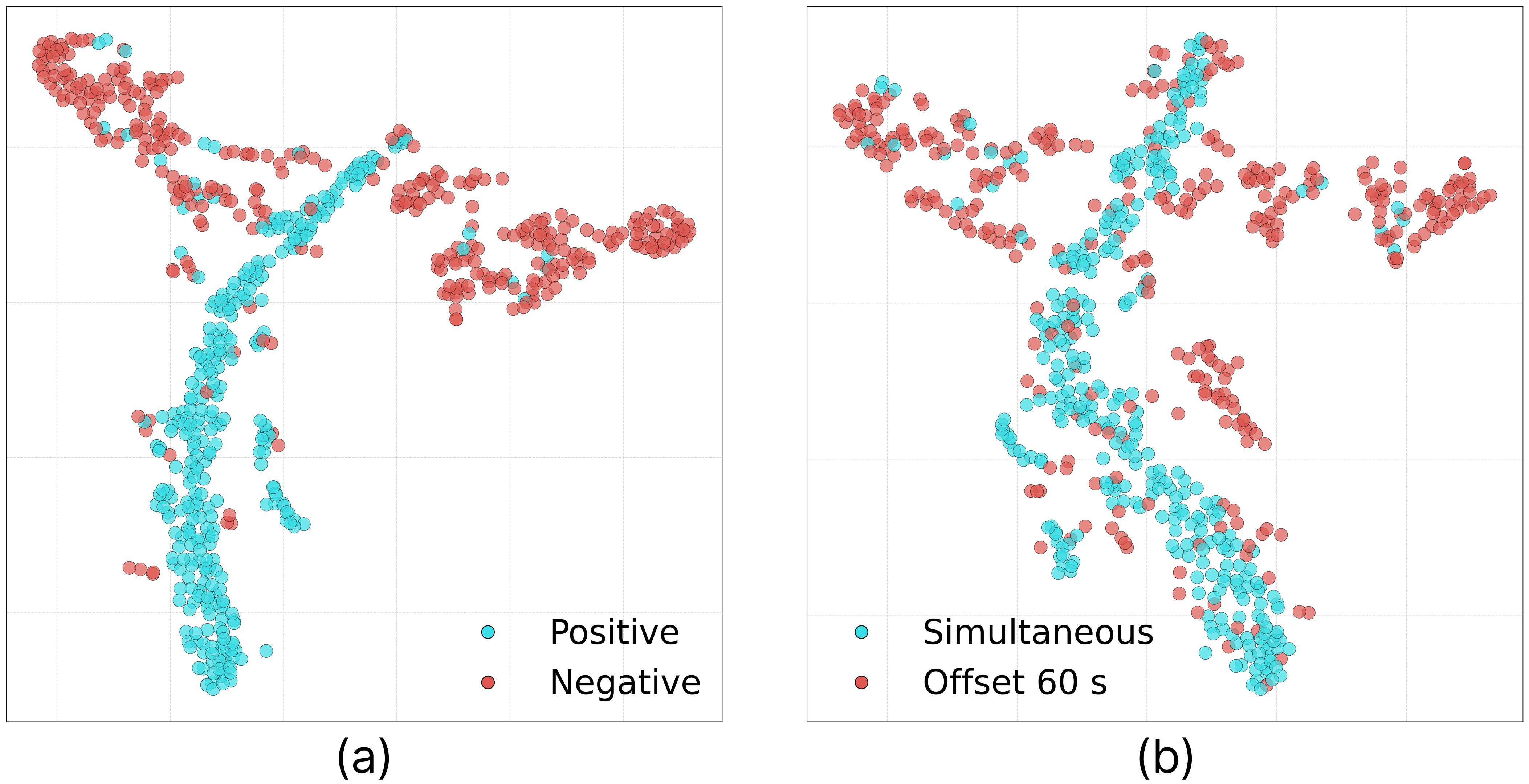}
  \caption{\textbf{t-SNE Visualization of Features.} Features from 5 random subjects before classifier implementation. (a) Blue: signals synchronously collected on different devices from the same subject. Red: signals from different subjects on different devices. (b) Blue: same as (a). Red: pairs from the same subject but with a 60s temporal offset.  
}  
  \Description{This figure presents two scatter plots visualizing t-SNE features from five subjects. In both plots, each point represents a feature pair. In (a), blue points show positive pairs—signals from the same subject collected at the same time—while red points show negative pairs from different subjects. In (b), blue points remain positive pairs, but red points represent signals from the same subject with a 60-second time offset. The visualization helps illustrate how feature distributions differ between subject identity and temporal conditions, supporting research on biometric signal classification.}
  \label{fig:T-SNE}
\end{figure}

We categorize signal segments into three classes: clean, weakly corrupted, and heavily corrupted. Segments with a relative spectral power score above 0.4 and a template matching score above 0.95 are classified as clean. The remaining segments are further evaluated using three criteria: (1) the presence of multiple pulse cycles with high correlation confidence, (2) temporal stability of inter-beat (RR) intervals, and (3) physiological plausibility of heart rate. A segment is labeled as weakly corrupted if it contains at least two high-confidence pulse cycles, exhibits low relative variability in RR intervals, and all RR intervals fall within a predefined physiological range. This rule-based screening retains segments with minor motion artifacts for mitigation using a specialized moving average filter~\cite{zhao2022robust}, while discarding heavily corrupted segments.


Signals passing MA processing are segmented into fixed 6-second windows. Each segment is standardized and smoothed with a 3rd-order Savitzky-Golay filter.

\subsection{Pairwise Feature Extraction}\label{sec:feature extraction}


\begin{figure*}[t]
    \centering
    \includegraphics[width=\linewidth]{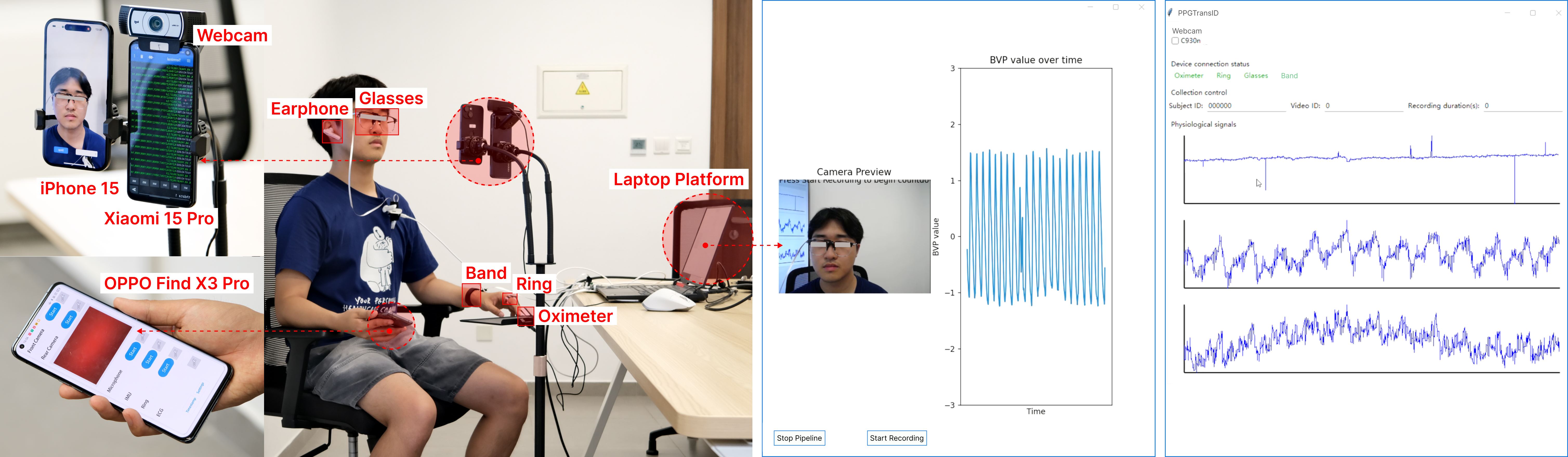}
    \caption{\textbf{Devices and Experimental Setup in Study 1.} Illustration of the experimental setup in Study~1, showing the wearing placements, device arrangements, and the platforms used on the smartphones and laptop.}
    \Description{This figure shows the experimental setup for Study 1, featuring a participant seated and wearing multiple devices. These include a custom-built ring, band, and glasses, as well as commercial products like an earphone, pulse oximeter, and several smartphones. The devices are positioned on the participant’s hand, wrist, face, and ear. The right side of the image displays a laptop interface collecting and visualizing physiological signals, such as blood volume pulse (BVP), from the devices. This setup demonstrates the integration of wearable technology for multimodal biosignal acquisition.}
    \label{fig:platform_setup_study1}
\end{figure*}



Signals are then paired according to study-specific rules. From each pair, 21 features are derived: 14 absolute differences (8 time-domain, 6 frequency-domain) and 7 similarity measures (6 time-domain, 1 frequency-domain). 
Full details of the extracted features are provided in Table~\ref{tab:ppg_features} in Appendix~\ref{appendix:features}.

Feature selection followed a two-stage process: importance ranking (via XGBoost, Random Forest, and SVM) and ablation analysis. Removing any selected feature consistently degraded performance (e.g., BAC dropped from 95.5\% to 95.3\% in XGBoost when the least important feature is discarded), confirming their necessity.

Figure~\ref{fig:T-SNE} visualizes the engineered features using t-SNE on 640 pairs from five randomly selected subjects. In (a), positive pairs (synchronous signals on different devices from the same subject) and negative pairs (different subjects and devices) form well-separated clusters, demonstrating strong discriminability. In (b), red points denote time-shifted pairs (60s offset) from the same subject, which deviate from simultaneous signals and align more closely with the negative class. This indicates that the system captures transient, time-varying features critical for CDA. Importantly, these plots only illustrate distribution trends of features before training; final performance is reported in Section~\ref{chap:study1}.

\subsection{Classification Algorithms}\label{sec:classifier}

We compared six classification models: three traditional machine learning baselines (XGBoost, Random Forest, and SVM) and three deep learning models (Siamese-LSTM, Siamese-ResNet18, and Siamese-RAPID~\cite{tang2025exploring}).

\subsubsection{Machine learning baselines.} XGBoost was trained with 100 estimators, a maximum depth of 6, and a learning rate of 0.1. Random Forest used 100 estimators with unrestricted depth, and SVM employed an RBF kernel with default parameters.

\subsubsection{Deep learning models.} For the deep learning approach, we adopted a Siamese architecture~\cite{varior2016gated}, which is well suited for comparison tasks such as authentication~\cite{zuo2018neural, wu2022use, ivanciu2021ecg, miller2021using, kim2025siamese}, and is effective even with relatively small datasets. Each network processes paired inputs through shared backbones, and the absolute difference between representations is fed into a fully connected layer for binary classification. We explored three backbone variants: \RI{(1) a two-layer LSTM with a hidden size of 512, (2) a standard 1D ResNet18 producing a 64-dimensional representation, and (3) RAPID~\cite{tang2025exploring}, a lightweight InceptionTime-style~\cite{ismail2020inceptiontime} architecture with attention mechanisms designed for PPG processing. The RAPID backbone in our implementation contains 18 RAPID blocks, while following the paper’s recommended kernel-set size of 9 and output dimensionality of 128.}

Prior to training, temporal difference and Welch transform were applied to generate three-channel inputs (original signal, temporal difference, and Welch spectrum). Models were optimized using binary cross-entropy with Adam (learning rate $1\times10^{-4}$), trained for up to 30 epochs with early stopping (patience = 5).

\section{STUDY 1}\label{chap:study1}



\RII{To address RQ1, RQ2, and RQ3,} we conducted a controlled lab study with 33 participants, collecting rPPG and PPG signals in two static postures.

\subsection{Platform Setup}\label{sec:study1 platform}

To support the evaluation study, we developed three data collection platforms: one for laptops and two for smartphones (iOS and Android), as shown in Figure~\ref{fig:platform_setup_study1}.

The laptop platform was implemented in Python with a Tkinter-based interface. It automatically discovered and connected to the custom wearable devices (Section~\ref{sec:hardware}). The platform visualized incoming PPG signals in real time. PPG signals, IMU data, and Unix timestamps were continuously logged during data collection.

For smartphones, we implemented two dedicated video recording platforms. The iOS application recorded facial videos using the front-facing camera, while the Android application recorded fingertip videos using the back camera. We also experimented with Android phones for facial video recording (tested on Xiaomi 15 Pro and OPPO Find X3 Pro). However, the Android devices apply non-configurable video compression, which significantly degraded the rPPG signal quality compared to iPhones. For both platforms, each video frame was timestamped and synchronized.

\subsection{Methodology}\label{sec:study1 methodology}

Data were collected in two static postures: sitting and standing, each lasting 10 minutes. We focused on these two conditions, as they represent the most common scenarios in which users interact with smartphones, aligning with the envisioned use cases of PPGTransID. Temperature and illuminance near the participant’s face were measured at the start of each experiment.

To synchronize timestamps across platforms (e.g., smartphones and the laptop), all devices were placed together in a box and subjected to three controlled drops prior to the experiment, allowing alignment via IMU peak detection. During data collection, an iPhone 15 and an external Logitech C930 webcam were positioned in front of participants for facial video capture. The ring, band, and oximeter were worn on the non-dominant hand, while the OPPO Find X3 Pro was held in the dominant hand, with its back camera covered by the participant’s fingertip.

Each session lasted approximately 35 minutes. Participants received compensation equivalent to 8.5 USD in local currency. The study protocol was reviewed and approved by the university’s Institutional Review Board (IRB).

\subsection{Demographics}\label{sec:study1 demo}

We recruited 33 participants (16 male, 17 female) with a mean age of 22.7 years (SD = 3.28, range = 18–30), 32 of whom were right-handed. The inclusion criterion was the absence of any diagnosed cardiac disease; no other physiological conditions were restricted.

\subsection{Evaluation Setup}\label{sec:study1 setup}

We adopted a leave-one-subject-out (LOSO) validation strategy, where models were trained on data from all but one participant and tested on the held-out participant.

rPPG signals were extracted from facial videos using both video-based and frame-based methods, and all PPG signals were then grouped into pairs for evaluation. Positive pairs consisted of signals from the same subject collected simultaneously on different devices, while negative pairs consisted of signals from different subjects and devices. Notably, non-simultaneous signals from the same subject were excluded from positive pairs (except in Section~\ref{sec:threat model II}). In testing pairs, one PPG signal was always taken from a token device (e.g., smartphone, laptop, or band; smartphone with the video-based rPPG extraction by default), while the other was taken from a wearable. For training, a 2-second overlap was applied for data augmentation.

To balance the dataset, the number of negative pairs was matched to that of positive pairs. Approximately 22 k pairs were generated for testing in the sitting scenario and 15 k in the standing scenario. The exact counts varied depending on the specific evaluation setup.

We adopt balanced accuracy (BAC) as the primary evaluation metric, defined as the average of the true positive rate and true negative rate. For each evaluation, we select the decision threshold that maximizes BAC. We additionally report the Area Under the Receiver Operating Characteristic Curve (AUC) and the Equal Error Rate (EER). AUC captures the overall discriminative capability of the system across thresholds, while EER corresponds to the operating point where the false acceptance rate equals the false rejection rate. An ideal authentication system achieves BAC = 1.0, AUC = 1.0, and EER = 0. All metrics are computed as weighted averages across subjects, with weights proportional to the number of test pairs per subject. 


\subsection{Performance of Different Methods}\label{sec:method comparison}

\begin{table*}[ht]
\centering
\caption{Different Models' Performance under Different Postures. Mean BAC of different methods under sitting (21,930 pairs) and standing (14,980 pairs) postures using LOSO validation on 33 subjects. XGBoost consistently achieved the highest accuracy across both conditions.}
\label{tab:method performance}
\begin{tabular}{llcccccc}
\toprule
\textbf{Method} & \textbf{Size} & \multicolumn{3}{c}{\textbf{Sitting}} & \multicolumn{3}{c}{\textbf{Standing}} \\
\cmidrule(lr){3-5} \cmidrule(lr){6-8}
& & EER$(\%)\downarrow$ & AUC$(\%)\uparrow$ & BAC$(\%)\uparrow$ & EER$(\%)\downarrow$ & AUC$(\%)\uparrow$ & BAC$(\%)\uparrow$ \\
\midrule
 Siamese-LSTM & 12.1M & 8.00 (5.61) & 96.2 (4.48) & 92.8 (5.46) & 10.3 (4.17) & 95.4 (2.53) & 90.7 (3.74) \\
 Siamese-ResNet18 & 14.8M & 8.10 (2.19) & 96.6 (1.37) & 92.8 (2.03) & 11.3 (3.74) & 94.9 (2.70) & 90.0 (3.40) \\
 Siamese-RAPID & 2.28M & 5.77 (1.77) & 98.0 (0.87) & 95.1 (1.63) & 8.85 (3.20) & 96.0 (1.97) & 92.3 (2.93) \\
 SVM & 16.0M & 9.24 (3.54) & 96.8 (1.63) & 91.7 (2.78) & 13.1 (5.22) & 94.0 (3.79) & 88.2 (4.28) \\
 Random Forest & 198M & 5.10 (1.94) & 98.6 (0.76) & 95.4 (1.81) & 7.93 (3.38) & 97.2 (2.03) & 92.8 (3.06) \\
 XGBoost & \textbf{0.46M} & \textbf{5.09 (1.86)} & \textbf{98.7 (0.73)} & \textbf{95.5 (1.66)} & \textbf{7.80 (3.41)} & \textbf{97.3 (1.93)} & \textbf{92.9 (3.05)} \\
\bottomrule
\end{tabular}
\end{table*}


\begin{figure}[t]
    \centering
    \includegraphics[width=0.4\linewidth]{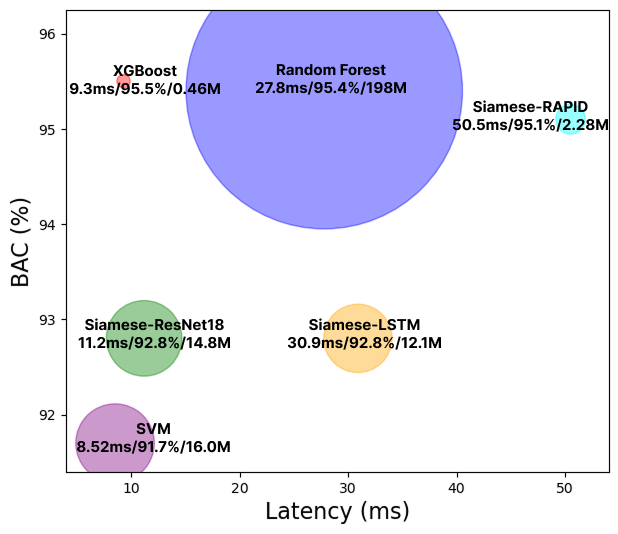}
    \caption{\textbf{Trade-off among Model Performance, Latency, and Size.} The y-axis shows mean BAC, the x-axis shows inference latency, and bubble size represents model size (number of parameters). 
    \textbf{XGBoost offered the best balance of accuracy, compactness, and speed.}}
    \Description{This bubble plot compares six models for biometric identification by showing their accuracy (BAC, \%) on the vertical axis, inference latency (milliseconds) on the horizontal axis, and model size as bubble area. Each bubble is labeled with its method name, latency, accuracy, and parameter count. XGBoost stands out for its high accuracy, low latency, and small size, while Random Forest has a large model size but similar accuracy. The plot highlights the trade-offs between speed, accuracy, and computational resources in model selection for biometric applications.}
    \label{fig:method latency}
\end{figure}

To compare the methods and assess PPGTransID's resistance against baseline attacks, we used rPPG signals extracted from iPhone 15 facial videos as the token device signals. Detailed results are reported in Table~\ref{tab:method performance}. We further benchmarked inference latency and model size (parameters) on an Intel Xeon Gold 5218 CPU and an NVIDIA RTX 3090 GPU. The results are summarized in the bubble plot in Figure~\ref{fig:method latency}. Notably, feature extraction time was included for traditional machine learning models (i.e., XGBoost, Random Forest, and SVM). 

XGBoost achieved the best overall trade-off, reaching the highest BAC in both postures (95.5\% sitting, 92.9\% standing) with only 0.46M parameters and a latency of 9.3 ms. Random Forest and Siamese-RAPID yielded comparable accuracy but required substantially higher complexity and slower inference. Siamese-LSTM and Siamese-ResNet18 delivered comparable BACs but required more memory and incurred longer latency. SVM lagged further behind with a $\sim$4\% drop while having a larger model size.

We attribute the lower performance of deep learning models to two factors: the relatively small dataset size and the simplicity of the features required for this task, which traditional models can capture more effectively. 

Given this balance of accuracy, memory efficiency, and speed, we adopt XGBoost as the default model in subsequent evaluations. Its performance demonstrates that \textbf{PPGTransID can effectively resist baseline attacks without subject-specific calibration (RQ2)}.

\subsection{Performance Across Devices}\label{sec:across device}

\begin{table*}[h]
\centering
\caption{Performance of video-based and frame-based rPPG extraction across four wearable devices in sitting and standing postures.}
\Description{This table compares authentication performance across different wearable devices and methods (video-based and frame-based rPPG) in sitting and standing conditions. Metrics include BAC, false positive rate, and false negative rate. High BAC values and low error rates for all devices indicate reliable and accurate identity verification. For blind and visually impaired users, these results support the system’s accessibility and effectiveness, enabling secure authentication with various wearables in diverse scenarios.}
\label{tab:device_performance}
\begin{tabular}{llcccccc}
\toprule
\textbf{Type} & \textbf{Device} & \multicolumn{3}{c}{\textbf{Sitting}} & \multicolumn{3}{c}{\textbf{Standing}} \\
\cmidrule(lr){3-5} \cmidrule(lr){6-8}
& & EER$(\%)\downarrow$ & AUC$(\%)\uparrow$ & BAC$(\%)\uparrow$ & EER$(\%)\downarrow$ & AUC$(\%)\uparrow$ & BAC$(\%)\uparrow$ \\
\midrule
\multirow{5}{*}{Video-based rPPG}
& Band      & 4.67 (2.39) & 98.7 (1,06) & 96.2 (2.09) & 9.01 (6.75) & 95.8 (5.65) & 92.5 (5.90) \\
& Ring      & 2.57 (2.01) & 99.4 (0.69) & 98.1 (1.53) & 6.18 (4.24) & 97.8 (2.10) & 95.2 (3.28) \\
& Glasses   & 6.99 (4.49) & 97.3 (2.84) & 94.4 (3.72) & 7.18 (7.13) & 96.7 (5.82) & 94.4 (4.83) \\
& Earphone  & 4.65 (2.91) & 98.6 (1.25) & 96.2 (2.34) & 6.55 (4.67) & 97.6 (3.12) & 94.6 (3.39) \\
& Total     & 5.09 (1.86) & 98.7 (0.73) & 95.5 (1.66) & 7.80 (3.41) & 97.3 (1.93) & 92.9 (3.05) \\
\midrule
\multirow{5}{*}{Frame-based rPPG}
& Band      & 5.86 (3.02) & 98.2 (1.59) & 95.2 (2.54) & 11.5 (8.23) & 94.4 (7.32) & 90.8 (7.01) \\
& Ring      & 3.64 (2.45) & 99.1 (3.31) & 97.2 (1.02) & 5.98 (3.37) & 97.7 (2.00) & 95.1 (2.65) \\
& Glasses   & 8.55 (5.39) & 96.3 (3.57) & 93.2 (4.28) & 11.7 (8.90) & 94.3 (7.60) & 90.9 (6.71) \\
& Earphone  & 6.18 (3.56) & 9.81 (2.45) & 95.4 (3.09) & 7.92 (4.34) & 97.0 (3.93) & 93.4 (4.00) \\
& Total     & 6.14 (2.82) & 98.1 (1.26) & 94.5 (2.60) & 9.07 (3.78) & 96.6 (2.17) & 91.7 (3.32) \\
\bottomrule
\end{tabular}
\end{table*}


\begin{figure}[t]
  \centering
  \includegraphics[width=1\columnwidth]{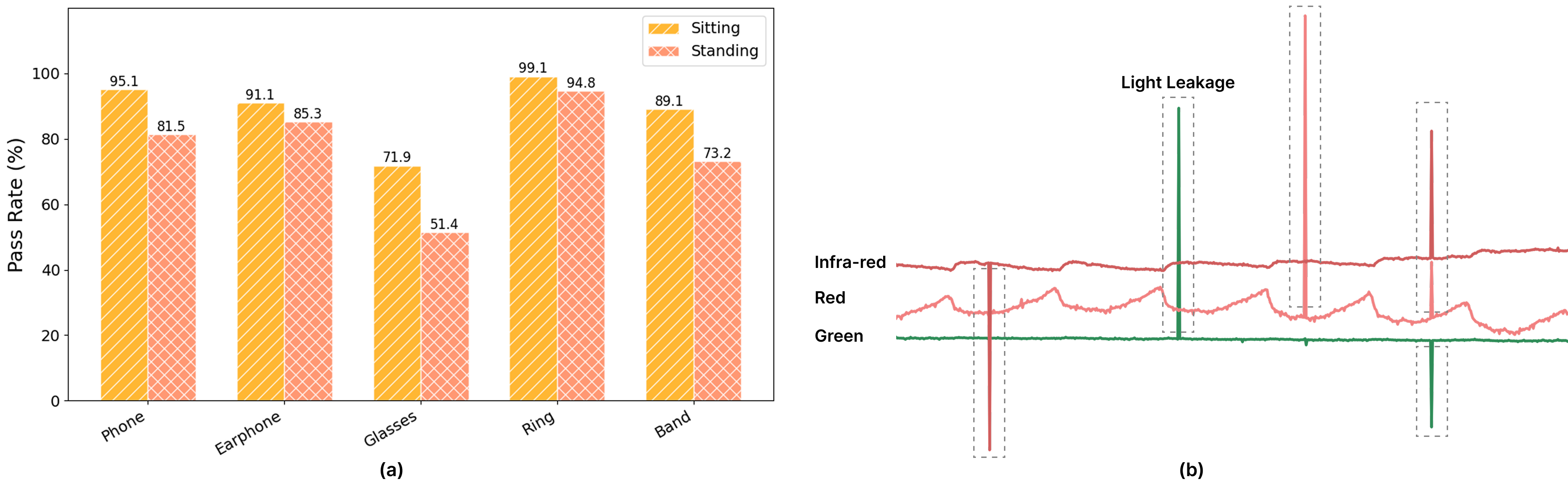}
  \caption{\textbf{Pass Rate across Wearables and Prototype Limitations.} (a) Pass rates for smartphone, earphone, glasses, ring, and band under sitting and standing postures, where ``Phone'' denotes video-based rPPG extraction. (b) Example of light leakage observed during data collection with the glasses prototype.}
  \Description{The bar chart shows pass rates for five devices under sitting and standing postures, with higher pass rates when sitting. The ring and smartphone achieve the highest pass rates, while the glasses show the lowest, especially when standing. The accompanying image illustrates light leakage observed in the glasses prototype during data collection.}
  \label{fig:pass rate}
\end{figure}

\RI{It is essential to evaluate whether the system remains reliable across different devices and to understand how user postures influence performance.} We formed evaluation pairs between each wearable device and the smartphone, reporting both authentication performance (Table~\ref{tab:device_performance}) and preprocessing pass rates (Figure~\ref{fig:pass rate} (a)).


All wearables achieve favorable accuracy with video-based rPPG extraction, with performance differences primarily driven by two factors. (1) PPG signal quality varies by body site: fingertip measurements yield the highest SNR, wrist signals are weaker and less stable, and ear-canal and nasal-bridge signals fall in between~\cite{tsai2021coherence, hartmann2019quantitative, budidha2013devepopment}. (2) The band, ring, and glasses are self-developed prototypes, and design constraints—such as imperfect skin contact, limited LED intensity, and light leakage—further affect signal quality. As shown in Figure~\ref{fig:pass rate} (b), severe light leakage in the glasses prototype results in a lower pass rate than other devices, which exceed 90\% accuracy in the sitting condition.



We also observed a consistent performance drop across devices in standing compared to sitting. This degradation is likely posture-induced: peripheral sensing sites, including the wrist, fingertip, ear canal, and nasal bridge, exhibit reduced PPG amplitude and SNR in the standing posture, consistent with decreased perfusion under hydrostatic pressure~\cite{linder2006using, yuan2017experimental, budidha2013devepopment}. The magnitude of this effect varies by device, with the band (–3.7\%) and ring (–2.9\%) showing the largest declines, consistent with their more distal vascular supply.



\RI{Overall, PPGTransID demonstrates robust cross-device performance with only modest degradation under standing posture (RQ3), supporting its applicability across diverse form factors (RQ1).}

\subsection{Influence of rPPG Extraction Methods}\label{sec:rPPG influence}

Table~\ref{tab:device_performance} compares the performance of two rPPG extraction methods. A consistent accuracy drop was observed when using the frame-based approach, reflecting the reduced rPPG quality. As discussed in Section~\ref{sec:rppg}, the video-based method leverages temporal information across the entire window, yielding higher-quality rPPG signals. In contrast, the frame-based method supports online inference but at the cost of accuracy. Despite this trade-off, both methods have distinct application potential: video-based extraction suits scenarios prioritizing reliability, while frame-based extraction enables low-latency, real-time deployment.

\subsection{Resistance Against Replay Attacks }\label{sec:threat model II}

\begin{figure}[t]
    \centering
    \includegraphics[width=0.5\linewidth]{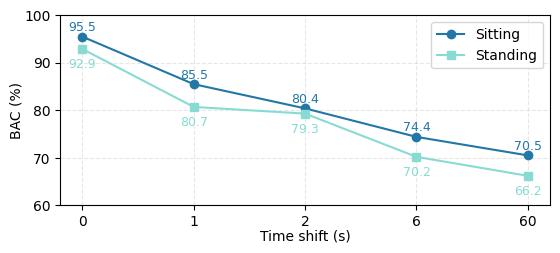}
    \caption{\textbf{Performance under Temporal Offsets.} BAC declines steadily with increasing temporal offsets between rPPG and PPG signals, highlighting the system’s reliance on instantaneous synchrony, indicating the system's resistance against replay attacks .}
    \Description{This line chart shows how biometric classification accuracy (BAC, \%) decreases as the time offset between paired rPPG and PPG signals increases. Two lines represent sitting and standing postures, both showing a steady decline in accuracy from nearly 95\% at zero offset to around 70\% at a 60-second offset. The results demonstrate that the system relies on signals being collected simultaneously, emphasizing the importance of temporal synchrony for reliable biometric identification.}
    \label{fig:time shift}
\end{figure}

Prior work has shown that PPG contains temporally stable features and can therefore support ODA across sessions; this raises the question of whether leaked historical PPG could threaten the proposed system. To evaluate this, we constructed positive pairs using rPPG and PPG signals from the same subject but recorded at different time intervals, simulating realistic delays in replay attacks. Results are summarized in Figure~\ref{fig:time shift}.

\RII{We observed a steady accuracy decline as temporal offsets increased. At a 60-second offset, performance dropped sharply to 70.4\% (sitting) and 66.2\% (standing), indicating that time-shifted pairs were increasingly misclassified as negatives.} This suggests the system relies on transient, synchronized features rather than invariant physiological patterns, making replay signals ineffective. 

In practice, replay attacks require multiple steps (e.g., interception, recording, injection), each introducing delays typically well above seconds to minutes. Our results show that such delays severely compromise attack feasibility. However, these results raise concerns about potential performance degradation caused by device-to-device latency. We examine this concern in Study~3 through a  Local Area Network (LAN)-based real-time demonstration.

In summary, these findings demonstrate that PPGTransID provides resistance against replay attacks (RQ2), leveraging transient PPG dynamics that are inherently difficult to capture or reproduce.

\subsection{Performance across Token Devices}\label{sec:token device}

\begin{figure}[t]
    \centering
    \includegraphics[width=0.4\linewidth]{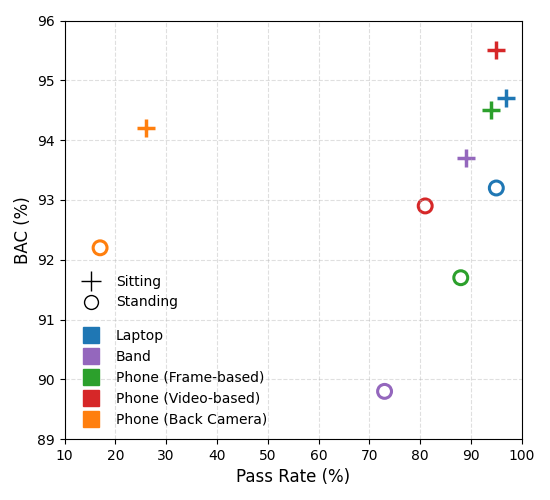}
    \caption{\textbf{Trade-off between Performance and Signal Pass Rate across Token Devices.} The y-axis shows BAC, while the x-axis indicates the pass rate after MA processing. Using the smartphone front camera for rPPG extraction achieves a favorable balance between accuracy, data utilization, and usability.}
    \Description{This scatter plot compares biometric accuracy (BAC, \%) and signal pass rate (\%) for different devices, including laptop, band, and smartphone using various camera modes. Each device is represented for both sitting (plus sign) and standing (circle) postures. Devices with higher pass rates generally achieve better accuracy, with the smartphone front camera showing the best balance of performance and data usability. The chart highlights how device choice and posture affect the effectiveness of biometric signal collection.}
    \label{fig:token device}
\end{figure}

Although smartphones were identified as the most suitable token devices in earlier discussions, we further examined the system’s applicability to alternative token devices. Following the same pairing procedure described in Section~\ref{sec:study1 setup}, the token device in the testing set was varied among a laptop, a band, the smartphone’s back camera (OPPO Find X3 Pro), and the smartphone’s front camera (iPhone 15). For the laptop, rPPG was extracted in real time using the frame-based method. \RII{For the back camera, rPPG was derived from fingertip videos~\cite{Lovisotto_2020_CVPR_Workshops}. Because token devices influence both classification accuracy and pass rate at the MA-processing stage (Section~\ref{sec:MA processing}), we report both metrics in Figure~\ref{fig:token device}.} 

Figure~\ref{fig:token device} shows that the system achieved consistent performance across token devices in the sitting posture, ranging from 93.7\% on the band to 95.5\% on the smartphone’s front camera. In the standing posture, performance remained acceptable on most devices, with the band again being the weakest (89.8\%). Device-specific differences reflect both signal quality and stability. The back camera suffered from a low pass rate due to unstable fingertip contact and light leakage. The front camera and the laptop showed stable performance, while the band was limited by lower signal quality.

Overall, while smartphones remain one of the most practical and effective token devices, our results confirm that PPGTransID generalizes well to alternative platforms. This flexibility enables broader deployment scenarios, such as laptop-based CDA when smartphones are unavailable.  

\RII{\subsection{Generalizability to Unseen Devices and Postures}\label{sec:unseen device}}

To examine the system’s ability to generalize to unseen devices (i.e., not included in the training set), we excluded each wearable device from the training set in turn and retrained the model without its data. The evaluation was then conducted on the same testing set used in Section~\ref{sec:across device} based on the video-based rPPG method.

\begin{figure}[t]
    \centering
    \includegraphics[width=0.6\linewidth]{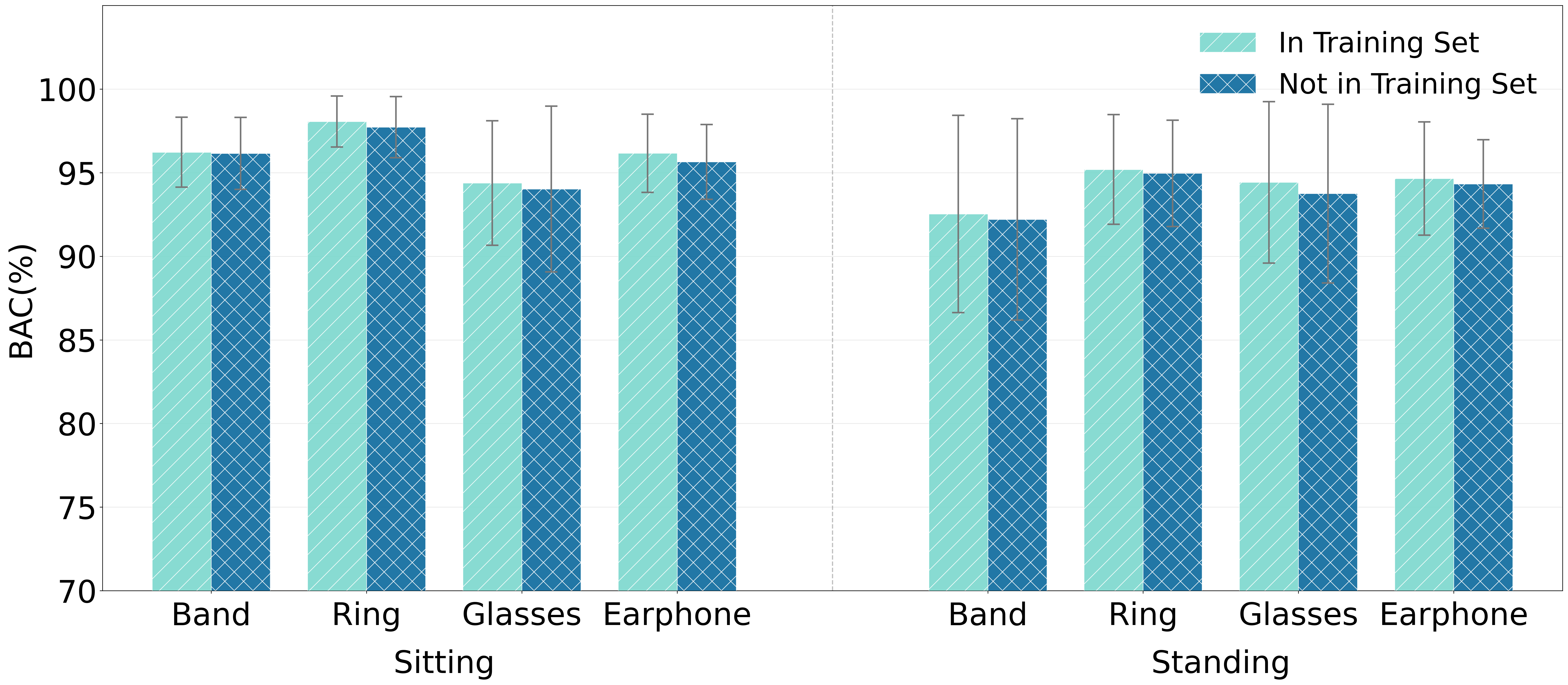}
    \caption{\textbf{Generalizability to Unseen Devices.} System performance when each wearable device is excluded from the training set. Experiments done under a strict cross-subject split. Whether each wearable device is in the training set has little impact on BAC, demonstrating generalizability to unseen devices and wearing locations.}
  \Description{This bar chart compares system performance (BAC\%) for four wearable devices—Band, Ring, Glasses, and Earphone—under sitting and standing conditions. It shows results for devices included and excluded from the training set. The similar BAC values and overlapping error bars indicate that the system generalizes well to unseen devices and wearing locations. This suggests the technology can reliably support blind and visually impaired users, regardless of which wearable device or position is used, enhancing accessibility and flexibility in real-world scenarios.}
    \label{fig:device generalizability}
\end{figure}

\RII{As shown in Figure~\ref{fig:device generalizability}, excluding any single device from training caused only marginal performance drops in both postures. The largest performance drop observed happens on earphone during sitting, decreasing from 96.2\% to 95.6\% BAC, which is negligible in practice and does not compromise system reliability.}

\RII{We also assessed generalizability to unseen postures by training exclusively on sitting data and testing on standing. Performance decreased slightly from 92.9\% to 92.3\% (0.6\% drop), indicating that posture changes minimally influence the system.}

\RII{Overall, these findings demonstrate that the system generalizes well to unseen wearables and postures without requiring fine-tuning, supporting its applicability across a broader range of real-world scenarios (RQ1).}


\subsection{Performance across Different Durations}\label{sec:duration}

The duration of facial video usage can significantly impact user experience. To evaluate the trade-offs, we varied the pre-processing window size and examined performance under shorter recording durations.

\begin{figure}[h]
    \centering
    \includegraphics[width=0.4\linewidth]{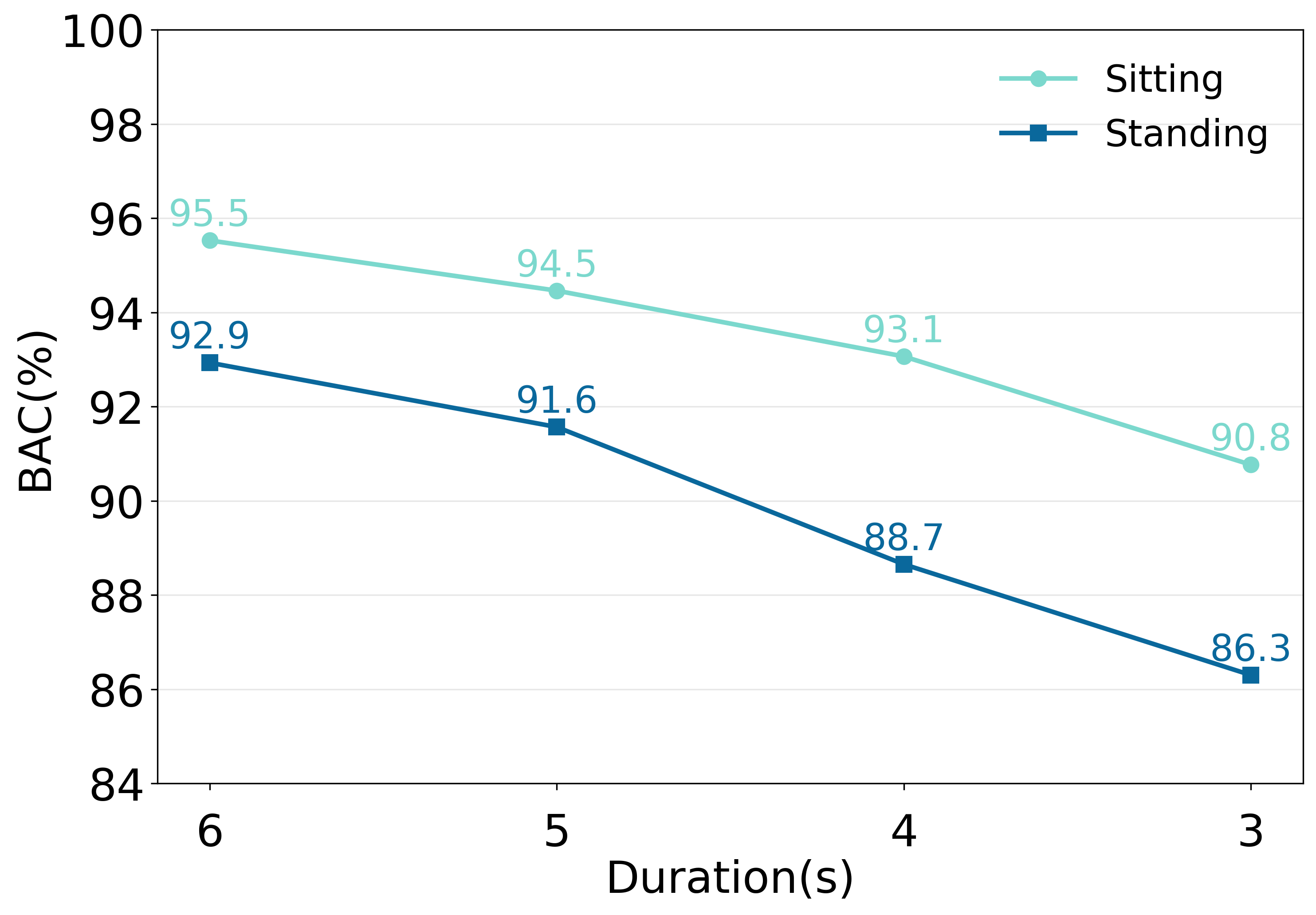}
    \caption{\textbf{Trade-off between Performance and Time Usage.} BAC steadily decreases as the time window duration becomes shorter, illustrating the trade-off between authentication accuracy and interaction efficiency.}
    \Description{This line chart shows how system performance (BAC\%) changes with different time window durations for sitting and standing conditions. As the duration decreases from 6 to 3 seconds, BAC drops, indicating a trade-off between authentication accuracy and speed. For blind and visually impaired users, this highlights the balance between quick, efficient interactions and reliable identity verification, helping to optimize accessibility without compromising security.}
    \label{fig:duration performance}
\end{figure}

As shown in Figure~\ref{fig:duration performance}, performance decreased with shorter windows: from 95.5\% (6s) to 90.8\% (3s) in sitting, and from 92.9\% to 86.3\% in standing. This highlights a clear trade-off between efficiency and accuracy: shorter windows allow faster interaction, whereas longer windows better support reliability-sensitive applications.  

The performance drop arises mainly from two factors. First, several features rely on stable statistics of the PPG signal, which become less reliable with shorter durations. Second, shorter windows capture fewer cardiac cycles, increasing susceptibility to anomalies and motion artifacts. Durations below 3 seconds were not tested, as they would yield fewer than five cardiac cycles—insufficient for feature extraction. While longer durations could improve accuracy, they demand extended stillness and reduce usable data, limiting practical deployment.

\subsection{Influence of Lighting and Temperature Conditions}\label{sec: study1 light condition}

Environmental factors such as lighting and temperature can affect rPPG extraction~\cite{yang2022assessment,tang2023mmpd,chen2024deep}. In our study, these conditions were not controlled: illuminance ranged from 270–3600 lx and temperature from 23.9–30.0 °C. To assess robustness, participants were grouped into three illuminance levels ($<700$ lx, 700–1100 lx, $>1100$ lx; $n=11,13,9$). A Kruskal–Wallis H test revealed no significant differences in BAC across illumination groups (H(2)=2.93, p=0.231), nor in rPPG preprocessing pass rates (H(2)=1.78, p=0.411). Participants were further divided into two temperature groups (below vs.\ above 26.5 °C). Mann–Whitney U tests again showed no significant differences in BAC (U=115, p=0.512) or pass rates (U=100, p=0.236).

Overall, these results indicate that typical variations in lighting and temperature do not significantly affect the reliability of PPGTransID (RQ3).

\subsection{Cross-session evaluation}

\IAC{Since CDA does not rely on template registration or long-term biometric stability, PPGTransID is not expected to experience cross-session degradation. The system operates purely on features extracted synchronously from co-present PPG signals and is trained in a user-independent manner.}

\IAC{To empirically validate this claim, we recalled 3 participants from the 33 participants after three months and collected new data following the same protocol. For each participant, we evaluated the new data using the model trained on the original 32 participants (excluding that participant’s own prior data). The average BACs before and after three months were 95.3\% (SD = 1.80\%) and 95.1\% (SD = 1.21\%), respectively, indicating comparable performance across sessions with no observable degradation.}

\IAC{These results confirm that, unlike ODA methods that depend on stable biometric templates, PPGTransID is inherently robust to cross-session variability.}

\begin{figure*}[t]
    \centering
    \includegraphics[width=1\linewidth]{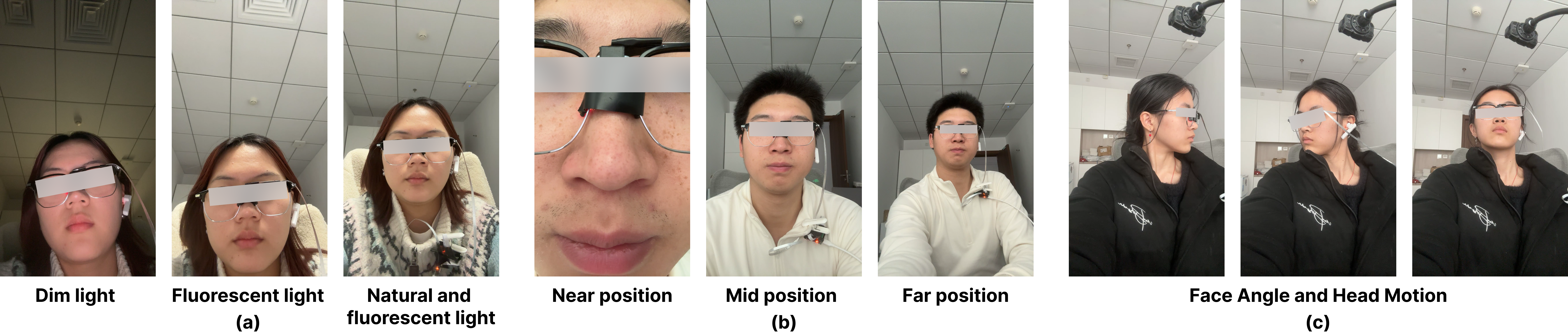}
    \caption{\IAC{\textbf{Examples of Different Lighting Conditions, Face Angles, and Camera Positionings.} (a) Front-camera frames under dim, fluorescent, and mixed natural–fluorescent lighting. (b) Frames captured at near, mid, and far camera distances. (c) Frames illustrating head motion at the far distance.}}
    \Description{This figure illustrates the experimental setup and visual conditions evaluated in Study 2. We systematically varied ambient illumination, participants’ head movements, and camera placement to assess their effects on PPG-based authentication performance.
(a) Example front-camera frames captured under dim, fluorescent, and mixed natural–fluorescent conditions.
(b) Example frames showing different camera positioning configurations and (c) head motion at the far position.}
    \label{fig:study3 methodology}
\end{figure*}

\IAC{\section{Study 2}\label{chap: study 3}}

\IAC{To further assess PPGTransID’s robustness to face--camera positioning, head movement, and lighting conditions (RQ3), we conducted a study with 10 participants under diverse settings.}

\IAC{\subsection{Methodology}\label{sec:study3 methodology}}


\IAC{Using the same platform and hardware setup described in Section~\ref{sec:study1 platform}, each participant wore all wearables and held an iPhone~15 to complete three recording sessions, each under a different lighting condition: (1) \textbf{dim lighting} with neither natural nor in-door illumination, (2) \textbf{fluorescent lighting} only with natural light blocked, and (3) \textbf{mixed natural–fluorescent lighting}. Example front-camera frames in each lighting conditions are provided in Figure~\ref{fig:study3 methodology} (a).}

\IAC{As shown in Figure~\ref{fig:study3 methodology} (b), for each lighting condition, participants recorded data at three distances. In the \textbf{near position}, the face was placed partially outside the camera frame to simulate partial facial visibility. In the \textbf{mid position}, the phone was held at a typical viewing distance during normal use. In the \textbf{Far position}, participants fully extended their arm and moved their head (up, down, left, right) to introduce different angles of the face.}

\IAC{At the beginning of each session, we recorded the illuminance level near the participant’s face.}

\IAC{Each session lasted approximately 30 minutes. Participants received compensation equivalent to 7~USD in local currency. The study protocol was reviewed and approved by the university’s Institutional Review Board (IRB).}

\IAC{\subsection{Demographics}}

\IAC{We recruited 10 participants (4 male, 6 female) with a mean age of 21.1 years (SD = 2.28, ranging from 18 to 25). None of the participants had taken part in Study 1.}

\IAC{\subsection{Influence of Lighting Conditions}\label{sec: study3 lighting}}

\IAC{To further examine how lighting affects the system’s usability and reliability, we evaluated PPGTransID under more extreme illumination levels than those in Section~\ref{sec: study1 light condition}, ranging from 18 lm to 36,800 lm. The average illuminance for the dim, fluorescent, and mixed natural–fluorescent lighting settings was 145 lm, 361 lm, and 5.68 klm, respectively.}

\IAC{For each condition, authentication pairs were constructed using the same procedure as in Study 1 (Section~\ref{sec:study1 setup}) and evaluated using the pretrained XGBoost model from Study 1 without any fine-tuning. The resulting BACs were 96.6\% (SD = 1.52\%), 96.6\% (SD = 1.69\%), and 96.1\% (SD = 2.74\%), with rPPG pass rates of 77.8\% (SD = 13.2\%), 83.8\% (SD = 11.7\%), and 70.8\% (SD = 25.4\%) for dim, fluorescent, and mixed lighting conditions, respectively. Although fluorescent light yielded a slightly higher pass rate, Kruskal–Wallis tests found no significant differences in either BAC (H(2)=2.93, p=0.231) or rPPG pass rate (H(2)=1.83, p=0.400) across conditions.}

\IAC{Overall, while extreme lighting can reduce rPPG pass rates, its impact on authentication accuracy is negligible, indicating that PPGTransID remains reliable across a wide range of illumination environments (RQ3).}

\begin{figure*}[t]
    \centering
    \includegraphics[width=\linewidth]{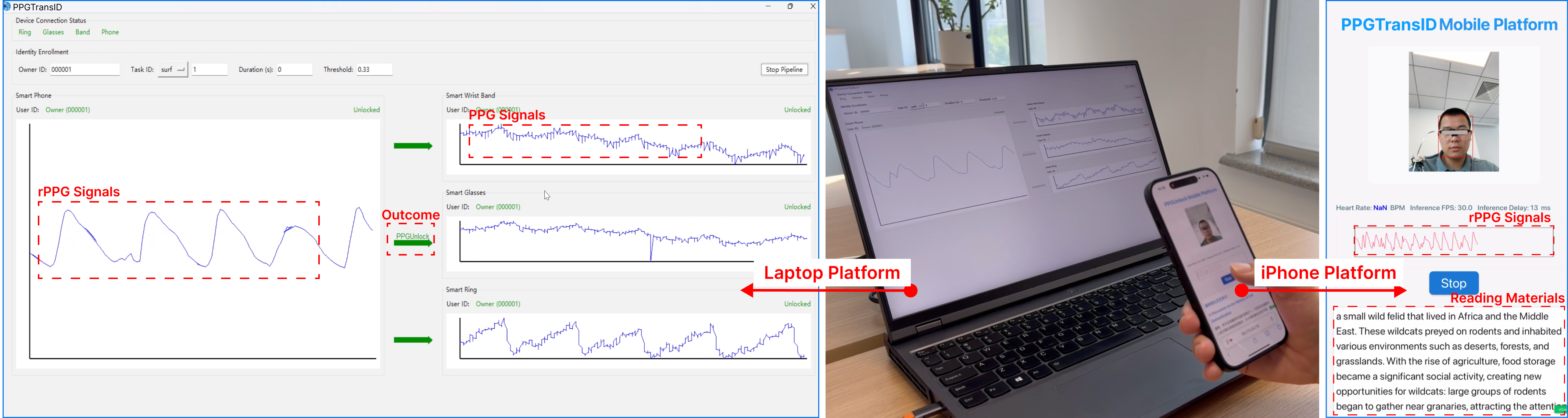}
    \caption{\textbf{Platform Setup in Study 3.} Two platforms (a laptop and an iPhone) were used for system demonstration and real-time performance evaluation. The iPhone runs a web app that extracts and transmits facial rPPG signals in real time to the laptop, which displays PPG/rPPG signals from multiple wearables along with the system interfaces.}
    \Description{This image shows the system setup using both a laptop and an iPhone. The iPhone runs a web app that extracts rPPG signals from facial videos and transmits them to the laptop. The laptop displays real-time PPG/rPPG signals from multiple wearable devices, such as a smartphone, wrist band, glasses, and ring. This multi-platform approach enables accessible, flexible identity verification for blind and visually impaired users, supporting various devices and usage scenarios.
}
    \label{fig:platform_setup}
\end{figure*}

\IAC{\subsection{Influence of Camera Positioning and Head Movement}\label{sec: study3 positioning}}

\IAC{rPPG extraction can be affected by partial facial visibility, extreme face–camera distances, and head movement, potentially impacting authentication reliability. To systematically assess these factors, we grouped them into three representative usage scenarios:
(1) \textbf{Near position:} the user holds a smartphone at an unusually short distance, causing partial facial occlusion;
(2) \textbf{Mid position:} typical smartphone usage distance;
(3) \textbf{Far position:} video-call–like scenarios with long face–camera distance and continuous head movement.}

\IAC{Using the same evaluation procedure as in the previous section, the resulting BACs were 97.5\% (SD = 2.05\%), 96.9\% (SD = 1.83\%), and 96.2\% (SD = 1.60\%) for the near, mid, and far conditions, respectively. Corresponding rPPG signal pass rates were 79.6\% (SD = 17.5\%), 85.8\% (SD = 15.4\%), and 66.8\% (SD = 21.8\%). Kruskal–Wallis tests indicate no significant differences across conditions for either BAC (H(2)=1.05, p=0.591) or pass rate (H(2)=4.25, p=0.120).}

\IAC{These results show that facial occlusion, suboptimal face–camera distance, and head movement can reduce signal pass rates due to degraded rPPG quality. Nevertheless, authentication accuracy remains stable, demonstrating that the system is robust to variations in camera positioning and moderate head motion (RQ3).}

\

\IAC{The evaluations demonstrate that PPGTransID maintains stable authentication performance across diverse lighting conditions, camera positions, and user behaviors—even in extreme scenarios (RQ3). While certain conditions may reduce rPPG signal pass rates, they do not compromise authentication accuracy. These results indicate that \textbf{PPGTransID is robust to real-world variability and holds strong potential for deployment in practical wearable authentication applications}.}

\IAC{}{\section{STUDY 3}}\label{chap:study2}

To address RQ2 and RQ4, we conducted a third study with 14 participants to evaluate the system's resistance to transmission delay and usability. Unlike Studies 1 and 2, which focused on offline evaluation with pre-collected data, this study emphasized live interactions and practical deployment contexts.

\subsection{Platform Setup}

To support this study, we developed two platforms—one on a laptop and one on an iPhone 15—to demonstrate the system and evaluate its performance in real-life scenarios, as shown in Figure~\ref{fig:platform_setup}.

The iPhone platform was implemented as a web-based application capable of capturing facial videos, extracting rPPG signals, and transmitting them to the laptop in real time. Signal transfer was handled through a self-hosted LAN server. To enhance user experience and improve system interpretability, the frame-based rPPG extraction method was adopted. The web interface included a live camera preview, a pipeline control button, and an AI-generated short article presented in both English and the local language.

The laptop platform was implemented in Python with a Tkinter interface, supporting rPPG reception, PPG collection, signal visualization, and system demonstration. Once the rPPG data were received from the smartphone, the system initiated a 6-second countdown for synchronized data collection. Synchronization was achieved by aligning the UNIX timestamp of the smartphone’s transmission with the laptop’s reception time. After each collection session, the platform automatically executed the proposed system and displayed the results: authenticated wearables were highlighted in green, failed authentications in red, and channels rejected by MA processing turned from light grey to dark grey. The system resets automatically after each evaluation.

\subsection{Methodology}\label{sec:methodology}

The study consisted of three sessions, each including three randomized tasks:  
(A) self-device authentication,  
(B) self-device authentication while scrolling, and
(C) baseline attacks.
Each task was repeated five times per session. The study included the three custom wearables (ring, band, and glasses). The earphone was excluded, as real-time signal access was not supported by the commercial device.

In Task A, participants wore the devices (ring and band on the non-dominant hand, iPhone 15 in the dominant hand) and initialized the pipeline by pressing the start button. They then recorded video and signals by looking at the phone in a natural and comfortable posture. Once the laptop displayed the authentication outcome, participants reviewed the result, after which the system reset automatically. Task B followed the same protocol, except participants read and scrolled through an article in a natural position at a normal pace during data collection. In Task C, participants acted as attackers, attempting to unlock the experimenter’s wearables. The procedure mirrored Task A, except the smartphone was controlled by the experimenter who acted as the owner.  


To account for potential variability, wearables were remounted after each session. After all three sessions, participants completed a post-study questionnaire on usability and perceptions. Each study lasted about 60 minutes, and participants received compensation equivalent to USD 14. The protocol was approved by the university’s IRB.

\subsection{Demographics}

We recruited 14 participants (9 male, 5 female) with a mean age of 24.1 years (SD = 3.48, range : 19–30), 13 of whom were right-handed. The inclusion criterion was the absence of any diagnosed cardiac disease; no other physiological characteristics were restricted. None of the participants had taken part in \RII{Studies 1 and 2}.

\subsection{Evaluation Setup}

\begin{figure}[t]
    \centering
    \includegraphics[width=0.6\linewidth]{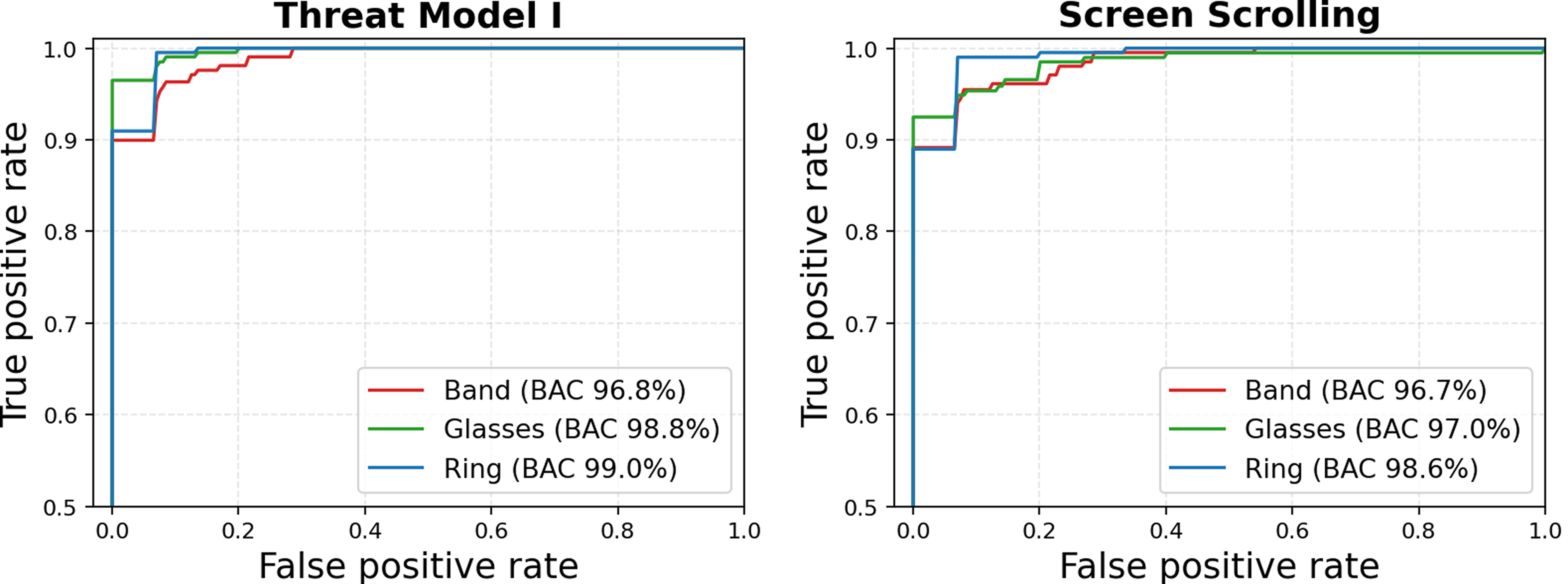}
    \caption{\textbf{ROC Curves for Wearables Under Three Experimental Conditions.} Averaged ROC curves for Band, Glasses, and Ring across grouped tasks. Some curves appear less smooth due to the restricted number of positive–negative pairs (30 pairs per subject and device).}
    \Description{This figure displays ROC curves for three wearable devices—Band, Glasses, and Ring—under different experimental conditions: Threat Model I and screen scrolling. Each device achieves high BAC values, demonstrating strong authentication performance. These results show that the system provides reliable and secure identity verification across various wearable devices and scenarios, supporting accessible and flexible use in daily life.}
    \label{fig:device roc}
\end{figure}

To assess system performance, we recorded the confidence scores of positive predictions generated by the XGBoost model trained \RII{in Study 1 without any fine-tuning. We then grouped task pairs (A+C and B+C) to calculate the average BAC across all 14 participants.} The results are summarized in Figure~\ref{fig:task roc}. In Figure~\ref{fig:device roc}, we also plotted the Receiver Operating Characteristic (ROC) curve for each wearable device for the grouped tasks.


\subsection{Real-time Resistance against Baseline Attacks} \label{sec:threat model I}

\begin{figure}[t]
    \centering
    \includegraphics[width=0.6\linewidth]{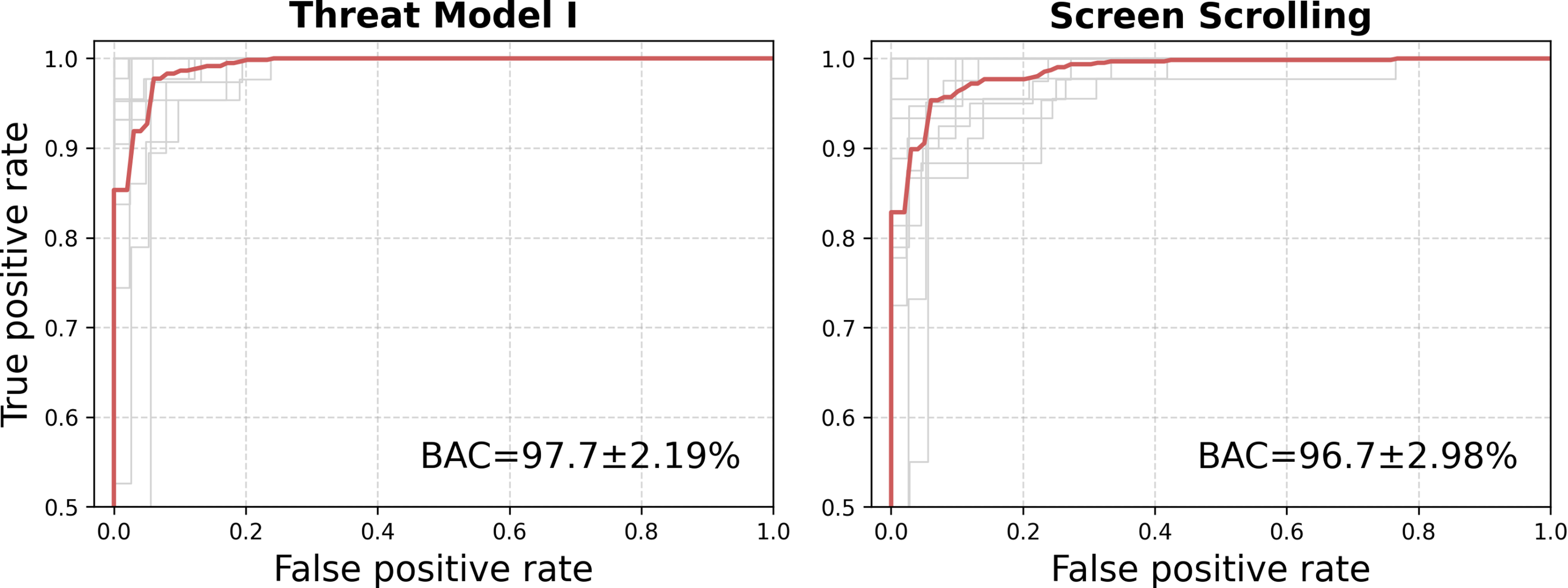}
    \caption{\textbf{ROC Curves for Three Experimental Conditions.} Each gray line shows the ROC curve of an individual participant, and the red curve represents the average ROC across all participants. The two panels correspond to distinct evaluation scenarios: Threat Model I (A+C) and CDA while scrolling (B+C).}
    \Description{This figure presents ROC curves for two experimental conditions: Threat Model I and screen scrolling. Each panel shows individual participant results (gray lines) and the average system performance (red line). High BAC values in all scenarios indicate strong and consistent authentication accuracy. This demonstrates the system’s reliability and robustness against different threats and usage conditions, supporting secure and accessible identity verification.}
    \label{fig:task roc}
\end{figure}

To evaluate the system's resistance against baseline attacks (Treat Model I) in a real-time demonstration, tasks A and C were grouped to calculate the metrics.

Figure~\ref{fig:task roc} shows that the system achieved a BAC of 97.7\%, outperforming the 95.5\% observed in Study 1. This increase was expected, as only two subjects were involved in forming negative pairs and the total number of tested pairs ($15,(repetitions)\times3,(devices)\times14,(subjects)\times2,(tasks)=1260$ pairs) was much smaller than in Study 1 (21,930 pairs), making the grouping problem relatively easier. Importantly, the system maintained stable performance under LAN-based data transmission latency, typically ranging from 5 to 20 ms~\cite{sui2016characterizing}, suggesting that such network delay does not hinder its functionality.


Overall, the results demonstrate that the system is resistant against baseline attacks, can sustain reliable performance in typical real-time usage scenarios, and is resilient to practical transmission delays.

\subsection{Generalizability on Screen Scrolling}

We envisioned that PPGTransID could capture facial videos from users in the background, without interrupting ongoing smartphone use. To evaluate its feasibility, tasks B and C were grouped to test the system's reliability under text-reading scenarios.

As shown in Figure~\ref{fig:task roc}, the system achieves a BAC of 96.7\%. Figure~\ref{fig:device roc} further shows that different devices obtain comparable performance. Among all devices, smart glasses exhibit the largest performance drop of 1.8\% compared to the sitting-still condition, indicating that the motion introduced by screen scrolling results in only a minor and acceptable degradation in performance. In practice, smartphone use frequently involves one-handed interaction~\cite{10.1145/3266037.3266129}, indicating that our system already covers a large portion of everyday use cases.


Taken together, these results demonstrate the practicality of PPGTransID during one-handed smartphone use. They show that rPPG signals can be reliably extracted from facial videos captured during natural interactions, enabling authentication with minimal interruption. This highlights the potential to unobtrusively integrate PPGTransID into mobile devices without disrupting user experience.

\subsection{Usability Results}

\begin{table}[h]
\centering
\caption{\textbf{Questionnaire dimensions and average scores.} Higher scores indicate more positive evaluation.}
\Description{This table summarizes questionnaire results on system usability, privacy, and acceptance. High average scores across dimensions such as ease of use, memorability, low physical demand, and attack resistance indicate positive user experiences. For blind and visually impaired users, these findings suggest the system is accessible, user-friendly, and secure, with strong social and privacy acceptance, supporting its potential for widespread adoption in assistive technology.}
\label{tab:short questionnaire}
\begin{tabular}{clc}
\toprule
\textbf{ID} & \textbf{Dimension} & \textbf{Average Score} \\
\midrule
Q1  & Smart wearable usage & 6.21 \\
Q2  & Privacy protection need & 6.43 \\
\hline
Q3  & Understandable & 6.35 \\
Q4  & Ease of use & 6.00 \\
Q5  & Authentication time acceptability & 5.29 \\
\hline
Q6 & Low cognitive load & 5.14 \\
Q7 & Memorability & 6.14 \\
Q8 & Low physical demand & 6.14 \\
Q9 & Low effort expended & 5.79 \\
Q10 & Low frustration & 5.64 \\
\hline
Q11 & Authentication accuracy & 5.79 \\
Q12 & Attack resistance & 6.14 \\
Q13 & Reliability & 5.93 \\
\hline
Q14 & Privacy acceptance & 5.86 \\
Q15 & Social acceptance & 5.86 \\
Q16 & Tendency to use & 5.79 \\
\hline
Q17 & Potential scalability & 6.00 \\
Q18 & Potential applications (open question) & N/A \\
Q19 & Open comments & N/A \\
\bottomrule
\end{tabular}
\end{table}

To evaluate the system’s usability, we administered a 7-point Likert-scale questionnaire to all participants at the end of the study. The evaluated dimensions and average scores are summarized in Table~\ref{tab:short questionnaire}, while the complete questionnaire can be found in Table~\ref{tab:full questionnaire} in Appendix~\ref{appendix:questionaire}.

\textbf{Usage and privacy concerns.} Participants first reported their frequency of using smart wearables (Q1, mean=6.21) and their approval of privacy protection on them (Q2, mean=6.43). These high scores indicate that the privacy concerns we aim to address are relevant to real users' needs.

\textbf{Explainability, user-friendliness, and efficiency.} The system was rated as highly explainable (Q3, mean=6.35) and user-friendly (Q4, mean=6.00). The authentication time was considered acceptable (Q5, mean=5.29), though some participants expressed concerns about efficiency. For example, P7 and P13 thought the process was slightly long, and P2 and P14 noted that “the system should be more efficient and work faster.” However, in our envisioned practical application, the system would run in the background rather than through the explicit interaction used in the study, which could improve user acceptance.

\textbf{Cognitive and physical demands.} Ratings for cognitive (Q6, mean=5.14), memory (Q7, mean=6.14), physical load (Q8, mean=6.14), effort (Q9, mean=5.79), and frustration (Q10, mean=5.64) suggest that most participants found the system lightweight to use. A few exceptions were noted: P2 and P13 reported higher cognitive load, and P9 and P14 mentioned frustration. These concerns were mainly attributed to the prototype UI rather than the system itself. For instance, P6 commented, “It would be better if the results could be displayed on the smartphone instead of the laptop.” Similarly, P14 suggested adding “an unlock animation to the UI, since it was tiring just staring at the screen.” We value those comments. As our UI served only as a proof-of-concept, we expect real-world deployments to make the process more seamless and implicit.

\textbf{Perceived reliability.} The system’s accuracy (Q11, mean=5.79), robustness (Q12, mean=6.14), and reliability (Q13, mean=5.93) were rated positively overall.Some participants voiced concerns about scalability to a larger scale of people (P7: “The system might fail as the subjects involved increase.”) and robustness after exercise (P12: “I’m worried the system might fail after exercise, since the heart rate is very different from resting.”). Our evaluation in Study 1 involved 33 participants with a LOSO validation, which partly addresses P7’s concern. Nevertheless, we did not examine robustness under post-exercise conditions, which remains an important direction for future work.

\textbf{Acceptance.} Finally, participants rated their acceptance of the system’s privacy usage (Q14, mean=5.86), its use in social contexts (Q15, mean=5.86), and their tendency to adopt it (Q16, mean=5.79). Most participants were positive, though P7 expressed privacy concerns and P13 reported reluctance to use it frequently.

\textbf{Scalability.} Participants were asked to rate the system’s scalability to potential use cases and to share their envisioned scenarios (Q17, mean=6.00). All participants agreed that the system holds strong potential for broader applications. The envisioned scenarios can be categorized into three main themes:

\begin{enumerate}
    \item \textbf{Security and access control.} Participants highlighted applications such as smart home (P10) and vehicle unlocking (P6, P7, P10), electronic door access (P2, P3, P9, P10), secure workplace entry (P9), banking multi-factor authentication (P6), and high-security environments requiring multiple biometric factors (P13). Some even envisioned forensic use cases, akin to fingerprint databases for law enforcement (P11).
    \item \textbf{Healthcare and wellness.} Several participants proposed medical applications, including automatic patient–record matching in hospitals (P3, P6), health data management (P1), and personalization in devices like smart scales (P7), where physiological identity could streamline individual data handling.
    \item \textbf{Personalization and privacy protection.} Participants also envisioned adaptive and privacy-preserving scenarios, such as identity-based car seat adjustment (P7), safeguarding personal data on wearable devices (P8, P12), and implicit continuous authentication across personal electronics (P12).
\end{enumerate}

Together, these envisioned scenarios extend beyond the CDA context and highlight a broader design space where physiological identity can be leveraged for both security and convenience. 

The valuable feedback above reflects users’ acceptance and confidence in different aspects of the system, supporting the system’s usability and its potential scalability to diverse use cases (RQ4).

\section{DISCUSSION}


This paper proposes PPGTransID, a ubiquitous method to enable cross-device authentication for smart wearables by leveraging the inherent consistency of PPG signals. Through the evaluation study (Section~\ref{chap:study1}), we demonstrate that PPGTransID is feasible across various wearables, generalizable to unseen devices and postures, and resistant against diverse threat models. In the real-time demonstration (Section~\ref{chap:study2}), users rated the system positively on usability and perceived reliability. Overall, we validated the ubiquity and usability of PPGTransID and envision the proposed CDA approach to facilitate smart wearable authentication in the future.





\subsection{Extending Strong Authentication to Lightweight Wearables}



The central idea of PPGTransID is to leverage smart devices with robust authentication mechanisms to empower lightweight wearable devices. Smartphones and laptops already provide mature solutions such as FaceID~\cite{schroff2015facenet} and fingerprint recognition~\cite{karu1996fingerprint}, whereas wearables face inherent constraints in space and computational power that limit direct deployment of comparable methods. PPGTransID bridges this gap by transferring authentication status from the former to the latter through a shared physiological signal. In this design, the smartphone functions as the trusted source device and effectively ``broadcasts" its identity via optical signals generated by blood volume changes. Furthermore, once authenticated, these wearables can themselves serve as token devices(Section~\ref{sec:token device}), creating a chained authentication mechanism that extends secure authentication throughout an ecosystem of devices. This not only ensures the trustworthiness of the token device but also enables a seamless extension of strong authentication from a phone or laptop to all associated wearables—embodying the principle of "authenticate once, use everywhere."
%

\subsection{Ubiquitous Cross-device Authentication via PPG}


A key strength of PPGTransID is its ubiquity. \RII{\textbf{PPGTransfer offers a unified CDA mechanism for common wearables without requiring independent ODA on each device.} This is feasible because PPG signals are widely accessible across today’s consumer devices.} Modern smartphones already integrate high-quality cameras capable of extracting rPPG~\cite{tang2023mmpd}, while most wearables—such as Whoop wristbands~\footnote{\url{https://www.whoop.com/}}, Oura rings~\footnote{\url{https://ouraring.com/}}, AirPods Pro earphones~\footnote{\url{https://www.apple.com/airpods-pro/}} and Meta Aria glasses~\footnote{\url{https://www.projectaria.com/glasses/}}—deploy optical PPG sensors for health monitoring. \RII{PPG signals extracted from these devices exhibit consistent cross-site patterns, enabling reliable CDA without requiring new hardware.} Our evaluation across four different devices (Table~\ref{tab:device_performance}) further demonstrates that this mechanism generalizes to diverse form factors. Beyond wearables, cameras and PPG sensors are increasingly embedded in emerging platforms such as smart mirrors~\cite{ma2025non} and automotive seats~\cite{wang2025physdrive}, suggesting that the same principle could extend to broader environments. \RII{\textbf{This ubiquity ensures that the system is not only deployable with current devices but also scalable to future ecosystems.}}

\subsection{Unobtrusive User Interaction}

A key advantage of PPGTransID is its unobtrusive user interaction. Prior CDA methods, such as ShakeUnlock~\cite{findling2016shakeunlock}, require explicit gestures like shaking or tapping to generate shared motion patterns. While effective for certain devices, these actions can be impractical or socially awkward, especially in public places where deliberate shaking may draw unwanted attention. Furthermore, motion-based methods are unsuitable for glasses and earphones worn on the head, and coordinating simultaneous gestures across multiple wearables can be cumbersome. In contrast, PPGTransID operates passively by leveraging physiological signals that are continuously collected during normal use. This design removes the need for deliberate actions, enabling users to unlock wearables easily while doing everyday activities on smartphones, such as screen scrolling or video watching—common smartphone uses that naturally place the face within view of the camera. The system can run quietly in the background during these everyday interactions, without requiring any extra effort from the user.


\section{LIMITATIONS AND FUTURE WORK}

\subsection{Duration Limitation of PPG Signal}
A limitation of PPG-based CDA lies in its reliance on cardiac cycles. Our method typically requires a short evaluation window of about 6 seconds (Section~\ref{sec:duration}), since multiple cardiac periods are needed to extract stable features. Unlike gesture-based CDA methods~\cite{findling2016shakeunlock}, which provide instant authentication, PPG-based transfer is better suited for implicit operation: the process can run unobtrusively in the background while the user interacts with their smartphone. However, in ultra-short verification scenarios (e.g., less than one second), the limited number of cardiac cycles constrains reliability. Future work may explore higher sampling rates, advanced modeling of partial waveforms, or predictive signal reconstruction to reduce the required duration and support faster interactions.

\subsection{Robustness to Strong Motion Artifacts}
Our studies focused on everyday postures such as sitting, standing, and screen scrolling, where PPGTransID performed reliably across conditions. However, scenarios with intense motion (e.g., running, cycling) were beyond the scope of this paper. PPG signals are highly susceptible to noise during vigorous activity~\cite{tang2023mmpd}, which could compromise accuracy. Addressing these contexts requires additional approaches, such as adaptive artifact suppression, multimodal fusion with IMU data, or generative methods that reconstruct clean signals from corrupted inputs. Extending the system into active settings would broaden its applicability to sports, rehabilitation, and mobile healthcare.

\subsection{Privacy Concerns for Using Physiological Signals}

Finally, the use of physiological signals raises potential privacy concerns. While our study shows that replaying leaked PPG data does not compromise security (Section~\ref{sec:threat model II}), users may still worry about both the act of camera capture and the potential exposure of sensitive physiological or emotional states, such as heart rate or stress levels~\cite{wang2024camera}. To mitigate these risks, our design requires explicit confirmation on the wearable before enabling camera recording, ensuring user awareness and control. While participants in our usability study reported high acceptance (Table~\ref{tab:short questionnaire}), future research should further investigate transparent consent mechanisms, on-device signal protection, and user controls over data retention. Balancing security and privacy will be critical for the responsible adoption of physiological CDA.

\IIAC{\subsection{Potential for Further Improving Security Performance}}

\IIAC{While PPGTransID achieves promising accuracy, this level may still be insufficient for security-critical applications. To further improve reliability, one avenue is to enhance MA mitigation using advanced PPG denoising models~\cite{Hung2025ReliablePM,afandizadeh2023accurate}, which can improve signal quality through modern deep neural network (DNN) approaches. A larger-scale data collection effort combined with contrastive pretraining could also further boost the performance of DNN methods. Finally, because our system is designed to run in the background with minimal interruption, continuous authentication is feasible and may help further raise the overall security level.}

\section{CONCLUSION}
This paper presented \textit{PPGTransID}, the first cross-device authentication system that leverages real-time consistency between rPPG from smartphones and contact PPG from wearables. PPGTransID enables a ubiquitous and unobtrusive authentication approach, allowing users to unlock wearables during routine smartphone use, such as text reading and browsing, without additional sensors or explicit gestures. Our evaluation demonstrated consistent performance across rings, bands, glasses, and earphones, achieving 95.5\% balanced accuracy in offline testing (N=33) and 97.7\% in real-time use (N=14), \IAC{while maintaining robustness to diverse environmental conditions and user behaviors and resisting both replay and mimicry attacks.} Participants further evaluated the system positively across key dimensions, including usability, reliability, privacy, and overall acceptance. Together, these findings establish PPG as a practical bridge for CDA in wearable ecosystems, opening new directions for secure and unobtrusive authentication.


\bibliographystyle{ACM-Reference-Format}
\input{main.bbl}

\appendix

\section{APPENDIX}

\subsection{Extracted Features}\label{appendix:features}


\RI{Table~\ref{tab:ppg_features} lists the 21 features extracted from each PPG pair, together with their full names and explanations. As shown in the table, we extract two categories of pairwise features:}

\RI{\textbf{Difference features:} For each characteristic $f_k$, we compute the absolute difference between the two signals}

\RI{\begin{equation}
D_k = \left| f_k^{(1)} - f_k^{(2)} \right|
\end{equation}}

\RI{where $f_k^{(1)}$ and $f_k^{(2)}$ are the values of the k-th feature extracted from two PPGs from a PPG pair, respectively.}

\RI{\textbf{Similarity features:} These directly measure how closely the signals co-vary, such as}

\RI{\begin{equation}
    S_k = \mathrm{sim_k}(x^{(1)}, x^{(2)})
\end{equation}}

\RI{where $x^{(1)}$ and $x^{(2)}$ denote the two PPGs from a pair and $sim_k$ is the k-th similariy feature.}



\begin{table*}[h]
  \caption{\textbf{Selected Features.} The 21 features extracted from each PPG-pair.}
  \Description{This table lists 21 features used to compare pairs of PPG signals, including time-domain and frequency-domain differences and similarities. These features capture heart rate, waveform shape, spectral properties, and signal correlations. For blind and visually impaired users, such detailed analysis enables accurate and secure identity verification using wearable devices, supporting accessible authentication without relying on visual cues.}
  \label{tab:ppg_features}
  \begin{tabular}{lll}
    \toprule
    Category & Name & Explanation \\
    \midrule
    Time-domain difference 
        & Heart rate & Mean heart rates \\
        & $PPI_{systolic}$ std. & Std. of systolic peak's pulse-to-pulse interval (PPI)~\cite{park2022photoplethysmogram}  \\
        & Mean pulse rise time (PRT) & Mean of the time from pulse onset to systolic peak~\cite{fine2022computational} \\
        & PRT std. & Standard deviation of PRT \\
        & PRT ratio & Mean PRT per cardiac cycle \\
        & PRT ratio std. & Std. of PRT ratio \\
        & Skewness & Distribution asymmetry of waveform \\
        & Kurtosis & Sharpness/peakedness of waveform distribution \\
    \midrule
    Frequency-domain difference 
        & Main frequency & Dominant spectral peak frequency \\
        & Second frequency & Secondary spectral peak frequency \\
        & High-frequency energy ratio & Energy ratio in band $>$1Hz \\
        & Low-frequency energy ratio & Energy ratio in band 0.5–1Hz \\
        & LF/HF ratio & Ratio of low- to high-frequency energy \\
        & Spectral entropy & Entropy of normalized power spectrum \\
    \midrule
    Time-domain similarity 
        & Coherence & Average coherence in 0.5–2Hz band \\
        & Maximum cross-correlation & Highest correlation coefficient \\
        & Maximum lag & Lag at which correlation peaks \\
        & DTW distance & Dynamic Time Warping distance \\
        & Pearson correlation & Linear correlation coefficient \\
        & Cosine similarity & Vector cosine similarity of signals \\
    \midrule
    Frequency-domain similarity 
        & Cosine similarity & Cosine similarity of power spectral densities \\
    \bottomrule
  \end{tabular}
\end{table*}

\subsection{User Study Questionaire}\label{appendix:questionaire}

Table~\ref{tab:full questionnaire} presents the exact questions that appeared on the questionnaire provided to the participants in Study 3.

\begin{table*}[h]
\centering
\caption{\textbf{Full Questionnaire Items.} Higher scores indicate more positive evaluation.}
\Description{This table lists all questionnaire items used to evaluate the system, covering smart wearable usage, privacy, ease of use, cognitive and physical demand, reliability, security, and social acceptance. For blind and visually impaired users, these questions assess whether the system is accessible, user-friendly, secure, and suitable for daily use, helping to ensure the technology meets their needs in real-world scenarios.}
\label{tab:full questionnaire}
\begin{tabular}{p{1cm}p{13cm}}
\toprule
\textbf{ID} & \textbf{Question} \\
\midrule
Q1  & I often use smart wearable devices such as smartwatches, bracelets, headphones, rings, glasses, etc. \\
Q2  & I think smart wearable devices need privacy protection. \\
\hline
Q3  & I think this system is easy to understand. \\
Q4  & I think this system is easy to use. \\
Q5  & I think the unlocking time required for this system is acceptable. \\
\hline
Q6 & I think using this system requires very little thought, attention, or cognitive processing. \\
Q7 & I think this system has a very low memory burden. \\
Q8 & I think using this system requires very little physical effort. \\
Q9 & I think using this system requires very little effort. \\
Q10 & I think using this system involves little frustration. \\
\hline
Q11 & I think this system can reliably unlock personal (wearable) devices. \\
Q12 & I think this system is very difficult to be effectively attacked. \\
Q13 & I believe this system can provide secure unlocking for smart wearable devices. \\
\hline
Q14 & I can accept the use of privacy in this system. \\
Q15 & I am willing to use this device in public places. \\
Q16 & I think I will use this system frequently. \\
\hline
Q17 & I think there are more potential application scenarios for this system. \\
Q18 & Possible potential application scenarios. (Open question) \\
Q19 & Suggestions for the system. (Open comments) \\
\bottomrule
\end{tabular}
\end{table*}

\subsection{LLM Usage Clarification
}

We used ChatGPT 5.0 to perform grammar check and correction. 

\end{document}